\documentclass{aa}  

\usepackage{graphicx}
\usepackage{xcolor}
\usepackage{booktabs}
\usepackage{supertabular}
\usepackage[version=4]{mhchem}
\usepackage{ulem}
\usepackage{longtable}
\usepackage{caption}
\usepackage{sidecap}
\usepackage{float}
\usepackage{placeins}
\usepackage{rotating}
\usepackage{txfonts}
\usepackage{hyperref}
\hypersetup{
    colorlinks=true,
    linkcolor= blue,
    filecolor=magenta,      
    urlcolor=blue,
    citecolor= blue,
}%

\begin{document} 

   \title{Hidden under a warm blanket: If planets existed in protostellar disks, they would hardly produce observable substructures}

   \author{P. Nazari,
          \inst{1}
          \and
          A. D. Sellek
          \inst{2}
          \and 
          G. P. Rosotti
          \inst{3}
          }

   \institute{European Southern Observatory, Karl-Schwarzschild-Strasse 2, 85748 Garching, Germany\\ 
        \email{Pooneh.Nazari@eso.org}
          \and
           Leiden Observatory, Leiden University, P.O. Box 9513, 2300 RA Leiden, the Netherlands
         \and
            Dipartimento di Fisica `Aldo Pontremoli', Universit\`{a} degli Studi di Milano, via G. Celoria 16, I-20133 Milano, Italy
             }

   \date{}

\abstract{The onset of planet formation is actively under debate. Recent mass measurements of disks around protostars suggest an early start of planet formation in the Class 0/I disks. However, dust substructures, one possible signature of forming planets, are rarely observed in the young Class 0/I disks, while they are ubiquitous in the mature Class II disks. It is not clear whether the lack of dust substructures in the Class 0/I disks indicates absence of planets or whether it is due to other physical effects such as temperature and dust opacity. Here we consider the effect of temperature on the ability of planets to produce dust substructures. We prescribe the evolution of the disk and the protostar from Class 0 to Class II phase and calculate the disk temperature using radiative transfer models at various stages of the evolution. We use the mid-plane temperature to calculate the disk scale height and the minimum planet mass needed to open observable dust gaps using the thermal criterion. We find that this minimum planet mass decreases as a function of time. Particularly, we find that if a planet up to ${\sim}5$\,M$_{\oplus}$ in the inner ${\sim}5$\,au or up to ${\sim}10-50$\,M$_{\oplus}$ at radii ${\gtrsim}5$\,au was already formed in the early protostellar phase ($t< 2\times 10^5$\,yr) it would barely produce any dust substructures. We conclude that a major contribution to the observed lack of substructures (if produced by planets) in the early protostellar phase - lowering their frequency by ${\sim}50\%$ - could be elevated temperatures rather than the absence of planets.}


   \titlerunning{If planets existed in protostellar disks, they would hardly produce observable substructures} 
   \maketitle
%

\section{Introduction}

The process of planet formation has been the subject of study for decades (\citealt{Lissauer1993}; \citealt{Pollack1996}; \citealt{Boss1997}; \citealt{Youdin2005}). Some of the most fundamental questions in this field relate to; when planets start to form, the typical timescales for planet formation, and the driving mechanisms behind planet formation. Considering the onset of planet(esimal) formation, given the long timescales of core accretion model (\citealt{Pollack1996}), it was long thought that they form in ``protoplanetary disks'' (i.e., Class II disks); once the envelope has dissipated, the disk is relatively mature, and lasts for longer (\citealt{Wuchterl2000}; \citealt{Goldreich2004}). However, this was questioned by recent Atacama Large Millimeter/submillimeter Array (ALMA) data finding that the masses inferred from millimeter continuum observations of Class II disks are at best equal to or at worst lower by one order of magnitude than the masses of exoplanets (\citealt{Najita2014}, \citealt{Manara2018}; \citealt{Tychoniec2018, Tychoniec2020}; \citealt{Tobin2020}).
Recent population synthesis work argues that the best match for the Class II dust mass distribution is indeed found when planets \citep{Savvidou2024}, or at least substructures \citep{Delussu2024}, have already formed early on.
Moreover, external photoevaporation (though maybe less relevant for disks in low-mass star forming regions such as the nearest and best-surveyed) can significantly limit disk lifetimes in the average stellar birth environment (\citealt{Winter2022}) and curtail the growth of planets \citep{Qiao2023}, nevertheless planets are ubiquitous in the Galaxy. This could imply that the formation of larger bodies - planetesimals or even planets - must start earlier in the Class 0/I stage while there is still enough mass to form them.

A popular mechanism for planetesimal formation is the Streaming Instability (SI) \citep{Youdin2005,Johansen2007,Johansen2014} which requires a sufficiently high concentration of solids with respect to the gas to trigger the clumping of solids with the clump eventually becoming self-gravitating. While some models predict that processes in the disk alone cannot achieve sufficient concentrations in the embedded stage \citep{Drazkowska2018}, the differential collapse of dust from the envelope with respect to the gas may enrich the protostellar disk enough to begin planetesimal formation (\citealt{Cridland2022}, see also \citealt{Tsukamoto2017}; \citealt{Koga2022}) and these disks may already have significant amounts of large dust grains and pebbles in their mid-plane (\citealt{Vorobyov2024}; see also \citealt{Bate2022}), although the high turbulence in young disks may impede planetesimal formation (\citealt{Lim2023}).
Once these first building blocks of planets form, they can then grow by pebble accretion, which is efficient due to the enhanced accretion cross-section because of the drag-assisted deflection and the steady supply of material from the drifting pebble flux \citep{Ormel2010,Lambrechts2012,Liu2020}. This process may be especially effective during the early stages of disk evolution due to the high accretion rates and pebble fluxes \citep{Tanaka2019}.
An alternative model of planet formation begins with the rapid fragmentation of the gas disk due to gravitational instability (GI) \citep{Boss1997}, which requires a high disk-to-protostellar mass ratio and thus is easier in young disks \citep[see review by][]{Kratter2016}. Drag forces may drive the instability on small lengthscales, resulting in the rapid formation of rocky cores of $1-10\,M_{\oplus}$ \citep{Longarini2023_theory,Longarini2023_sim,Baehr2023} during the protostellar phase.

Nevertheless, the early formation of planets is still actively debated and observers are searching for more evidence of this phenomenon in Class 0/I stages. Particularly, if planets truly start to form early, it might be expected to observe their dust substructure signatures in these early stages similar to those commonly observed in mature Class II disks (\citealt{Calvet2002}; \citealt{Andrews2016}; \citealt{Akiyama2016}; \citealt{Fedele2017}; \citealt{Pinilla2018}; \citealt{Francis2020}; \citealt{Long2023}). Multiple observational studies searched for these substructures, and although some works find a few systems with dust rings and gaps in young protostellar disks (\citealt{Sheehan2018,Sheehan2020}; \citealt{Segura2020}), the majority of Class 0/I disks do not show any substructures (see the review by \citealt{Tobin2024}). In particular, recent statistical studies of protostellar disks such as the Early Planet Formation in Embedded Disks (eDisk) ALMA large program and the ALMA Legacy survey of Class 0/I disks in Corona australis, Aquila, chaMaeleon, oPhiuchus north, Ophiuchus, Serpens (CAMPOS) considered 19 and 124 Class 0 and I disks at ${\sim}0.04\arcsec$ and ${\sim}0.1\arcsec$ resolution, respectively. Both surveys had a substructure detection rate of ${\lesssim}15\%$ (\citealt{Ohashi2023}; \citealt{Hsieh2024}), while this number is ${\gtrsim} 50\%$ in Class II disks (\citealt{Baruteau2014}; \citealt{Andrews2018}; \citealt{Long2018}; \citealt{Cieza2021}; \citealt{Bae2023}). These numbers could be higher if we account for some resolution bias, but not all substructures are due to planets (e.g., \citealt{Zhang2015}; \citealt{Flock2015}; \citealt{Takahashi2016}; \citealt{Tzouvanou2023}), and some planets might open multiple rings and gaps (\citealt{Bae2017, Meru2019}), suggesting massive planets in embedded disks may be quite rare.

Therefore, considering the expected timescales for planet formation (e.g., \citealt{Lambrechts2014b, Drkazkowska2023, Savvidou2023, Lau2024}), a relevant question is whether these relatively smooth Class 0/I disks indeed imply the absence of planets altogether, or are simply the result of other physical effects such as temperature and opacity effects. It has been suggested that the lack of substructures could be due to dust opacity effects (\citealt{Ohashi2023}; \citealt{Sharma2023}; \citealt{Guerra-Alvarado2024}). In this work, we consider the effect of temperature on the ability of planets to carve a gap - and the time of gap opening - if they were to exist in embedded disks.

Observations find that young Class 0 and I disks are in general warmer than the Class II disks (\citealt{vantHoff2020}; \citealt{Takakuwa2024}). This is also implied from observations of highly complex millimeter spectra toward these objects where molecular species are sublimated from ices into the gas in warmer regions as opposed to the nearly barren spectra of Class II disks (\citealt{Thi2004}; \citealt{Jorgensen2016}; \citealt{Belloche2020}; \citealt{Loomis2020}; \citealt{Yang2021}; \citealt{Nazari2024}). The higher temperatures are expected to result in larger disk scale heights ($H/r \propto \sqrt{T}$), which in turn make it more difficult for planets to carve a gap (\citealt{Lin1993}; \citealt{Crida2006}; \citealt{Baruteau2014}) with the pebble isolation mass therefore a declining function of time (\citealt{Bitsch2015b}). In this work, we use radiative transfer models to calculate the temperatures and thus the scale heights of disks over time starting with early Class 0 and end with Class II disks. We then use analytical relations from hydrodynamical simulations to relate this to the minimum planet mass needed to carve a gap ($M_{\rm gap}$) as a function of disk evolution. We find that this minimum mass decreases as a function of time and even if a gas giant accreted ${\sim} 20\%$ of its mass in the protostellar phase, it would likely not open a gap. These findings have important implications for planet formation timescales which we consider further in the discussion. 

\section{Models}

Here we simulate the evolution of the disk around a protostar from Class 0 phase to Class II phase by combining the models of \cite{Fischer2017} and pre-main sequence evolutionary tracks of \cite{Baraffe2015}. The detail of our evolutionary model is given in Appendix \ref{app:model} and Fig. \ref{fig:evolution} shows the prescribed evolution of various parameters in our model. The prescribed values qualitatively match the observational measurements and other evolutionary models available in the literature (e.g., \citealt{Evans2009}; \citealt{Visser2009}; \citealt{Dunham2010}; \citealt{Kristensen2012}; \citealt{Pokhrel2023}). 

\begin{figure}
    \centering
    \includegraphics[width=0.85\columnwidth]{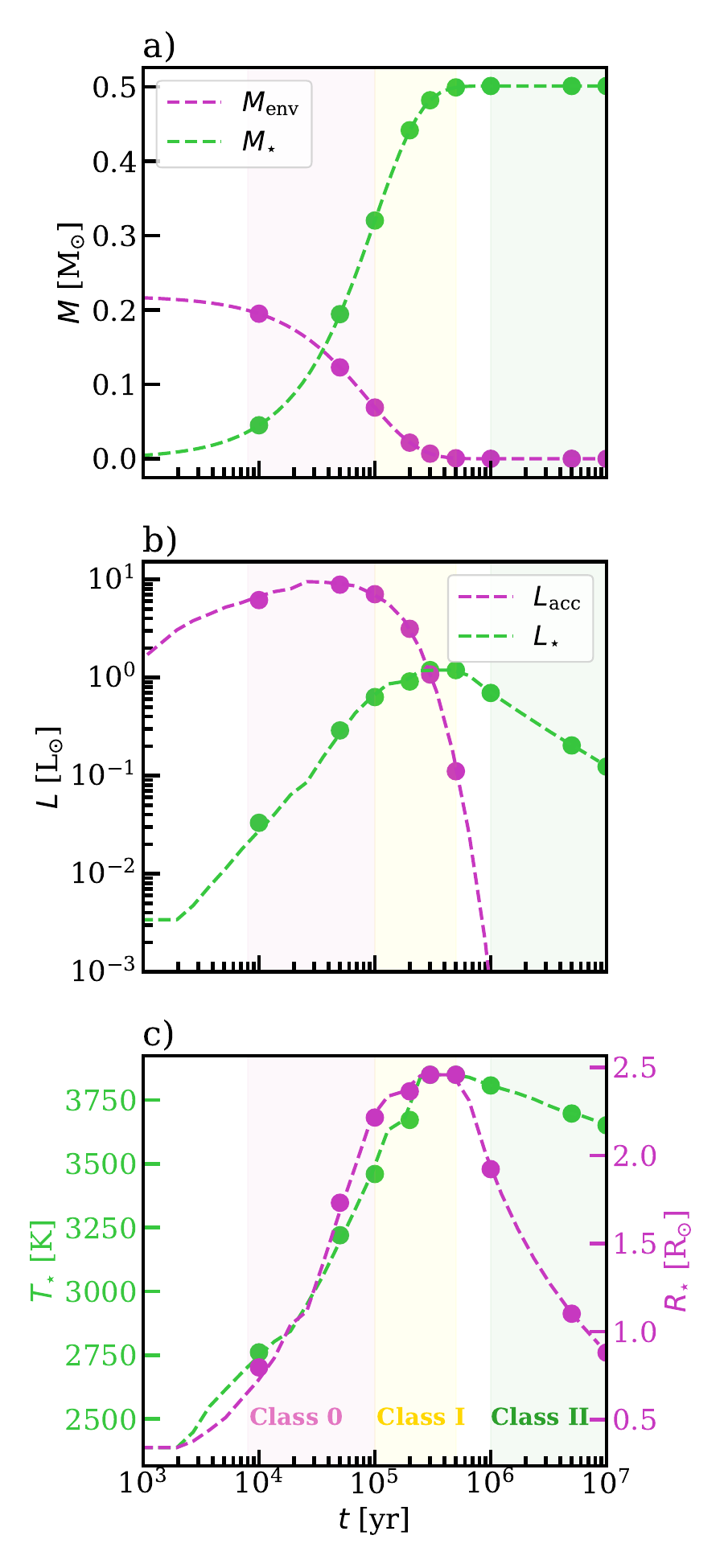}
    \caption{The evolution of various parameters as the protostar ages for a final protostellar mass of 0.5\,M$_{\odot}$. a) Evolution of envelope and protostellar mass. b) Evolution of accretion and protostellar luminosity. c) Evolution of protostellar radius and effective temperature. The filled circles indicate the nine distinct times that we have chosen to run the radiative transfer models. Pink, yellow and green shaded areas roughly indicate the different Classes.}
    \label{fig:evolution}
    \vspace{-12pt}
\end{figure}

We then calculate the temperature using radiative transfer at nine distinct times during the disk evolution to find the disk scale height as a function of evolution. The radiative transfer models follow the general methods of \cite{Nazari2022_disk}. We use RADMC-3D version 2.0\footnote{\url{http://www.ita.uni-heidelberg.de/~dullemond/software/radmc-3d}} (\citealt{Dullemond2012}) for our calculations. The density structures of the disk and the flattened envelope are the same as disk-plus-envelope models of \cite{Nazari2022_disk}. The disk gas surface density was $\propto 1/R$, where $R$ is the radius in cylindrical coordinates. The assumed parameters of the models for various $M_{\star, \rm f}$ at different evolutionary ages are given in Tables \ref{tab:params_0.1}-\ref{tab:params_1}. For more information on our choices for the final parameters used in the radiative transfer models, see Appendix \ref{app:params}.

The major difference between our models and those of \cite{Nazari2022_disk} is the iterative nature of our procedure to calculate the temperature. We initially start with $\epsilon$ values of 0.3, 0.2, and 0.1 for $t < 10^{5}$\,yr, $10^5\leq t < 10^{6}$\,yr, and $t \geq 10^6$\,yr, respectively. The $\epsilon$ value was used in \cite{Nazari2022_disk} to prescribe the disk scale height as $H(R) = \epsilon R$. The initial values of $\epsilon$ are based on observational measurements of disk scale heights in various stages of a disk (\citealt{Pinte2018}; \citealt{vantHoff2020}; \citealt{Law2022}; \citealt{Paneque2022}; \citealt{Lee2022}). Once we calculate the disk temperature using this initial scale height prescription, we re-calculate the disk scale height using the calculated disk mid-plane temperature using (e.g., \citealt{Armitage2010})

\begin{equation}
    H(R) = c_{\rm s}/\Omega = \sqrt{\frac{k_{\rm B}T_{\rm mid-plane}R^3}{\mu m_{\rm H} G M_{\star}}},
    \label{eq:H_r}
\end{equation}

\noindent where $c_{\rm s}$ and $\Omega$ are the sound speed and angular velocity. In Eq. \eqref{eq:H_r}, $k_{\rm B}$, $T_{\rm mid-plane}$, $m_{\rm H}$, and $G$ are Boltzmann constant, mid-plane temperature, mass of hydrogen and gravitational constant, respectively. We first smooth the calculated mid-plane temperature using \texttt{gaussian\_filter} from \texttt{scipy.ndimage} in \texttt{python} before calculating $H(R)$ using Eq. \eqref{eq:H_r}. We repeat the above procedure for two more iterations where various tests showed that it is sufficient for the mid-plane temperature to converge within ${\sim}15\%$ and for most models within ${\sim}5\%$ at radii of ${\gtrsim}10$\,au which are the focus of this work. We note that our conclusions are independent of the initial $\epsilon$ assumed. We use 300 grid points for $r$ between 0.4 and 0.5\,au and use 700 grid points for $r$ between 0.5 and 150\,au. We use 200 grid points in the $\theta$ direction between 0 and $\pi/2$, where $r$ and $\theta$ are the radius and polar angle in the spherical coordinates. Moreover, we use three million photons to calculate the temperature initially and then six million and 10 million photons for the next two iterations. These values were chosen based on various tests to ensure a robust measurement of the temperature.

\section{Aspect ratio and minimum gap-opening mass}
\label{sec:results}

\begin{figure*}
    \centering
    \includegraphics[width=0.9\textwidth]{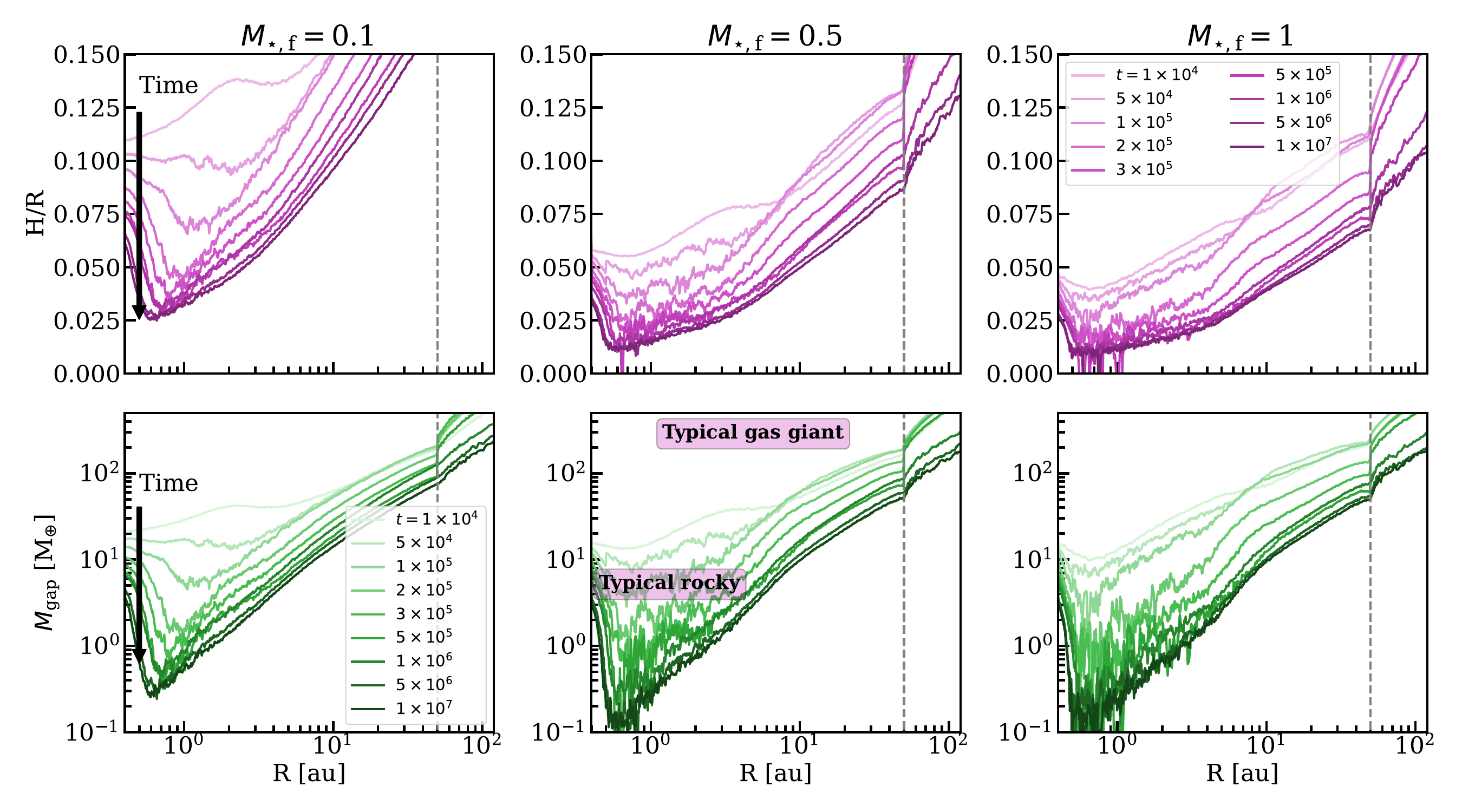}
    \caption{The evolution of disk aspect ratio (top) and minimum planet mass required to open a gap (bottom) from young Class 0 disks (lighter colors) to mature Class II disks (darker colors) for different final protostellar masses. These models are those with small dust grains (0.1\,$\mu$m). The gray dashed line presents the radius of the disk assumed in these models. The typical masses of giant and rocky planets are taken from \cite{Zhu2021} (also see \citealt{Akeson2013}). Up to ${\sim}50$\,M$_{\oplus}$ of a giant planet and an entire rocky planet would create no sign of a dust gap if formed in the protostellar phase ($t< 2 \times 10^{5}$\,yr).} 
    \label{fig:min_pl_mass}
\end{figure*}

The aspect ratio ($H/R$) calculated from the mid-plane temperature using Eq. \eqref{eq:H_r} for disk scale height is presented in the top row of Fig. \ref{fig:min_pl_mass}. This figure shows that there is a divide between times ${\lesssim} 10^5$\,yr and ${\gtrsim} 5 \times 10^5$\,yr. At 1\,au, the aspect ratio at early times is ${\gtrsim}0.05$ higher than late times for $M_{\star, \rm f} = 0.1$\,M$_{\odot}$. This difference is ${\gtrsim}0.025$ for $M_{\star, \rm f} = 0.5$\,M$_{\odot}$ and 1\,M$_{\odot}$. At 10\,au the early times are ${\gtrsim}0.05$ higher than the late times for all final protostellar masses. This large drop in the aspect ratio is driven by the final accretion of the envelope onto the star+disk, which leads to a large drop in the accretion luminosity and thus the irradiation of the disc. A slower decline in the aspect ratio on timescales ${\gtrsim}1\,\mathrm{Myr}$ is driven by the luminosity falling as the star contracts down the Hayashi track. The disk aspect ratio for models with a disk radius of 100\,au (Fig. \ref{fig:min_pl_mass_100au}) and models with large dust grains (Fig. \ref{fig:min_pl_mass_large}) have a similar trend. That is, they generally get smaller for older disks. The results of these models are discussed further in Appendix \ref{app:other_models}.

The minimum planet mass required to open a gap in a disk ($M_{\rm gap}$) has been studied with hydrodynamical simulations of Class II disks (\citealt{Lambrechts2014a,Rosotti2016,Bitsch2018}), which typically show that the thermal criterion - that the Hill radius of the planet must reach of the order of the scale height \citep{Lin1993,Paardekooper2006, Baruteau2016, Ataiee2018} - controls the onset of gap opening. Quantitatively, the gap becomes observable in mm observations tracing large dust grains slightly before this point, once the mass of the protoplanet reaches approximately $20\,M_{\oplus}$ for a solar-mass protostar \citep{Rosotti2016}. The required protoplanet mass has a linear scaling with protostellar mass at fixed aspect ratio and typical (using $\alpha$ prescription of \citealt{Shakura1973} with $\alpha = 10^{-3}$) disk viscosities (\citealt{Liu2019, Sinclair2020}, although the weaker gravitational potential of low-mass stars results in their disks being thicker and thus having a higher gap-opening mass):

\begin{equation}
    M_{\rm gap} \simeq 20\,M_{\oplus} \frac{M_{\star}}{M_{\odot}} \left(\frac{H/R}{0.05}\right)^3.
    \label{eq:min_pl_mass}
\end{equation}

\noindent We emphasize that this relationship is based on hydrodynamical simulations of Class II disks where no envelope is present. Therefore, this relation only gives a lower limit on the minimum planet mass needed to \textit{observe} a gap, since in the Class 0/I stage the envelope and dust extinction can hinder the observation of the gap. Moreover a thicker disk geometry can obfuscate any gaps that do open when viewed at high enough inclination \citep{Guerra-Alvarado2024}. In other words, the minimum planet masses needed to observe a gap could be even higher for Class 0/I disks than given here. 

Using Eq. \eqref{eq:min_pl_mass}, $M_{\rm gap}$ for our models is presented in the bottom row of Fig. \ref{fig:min_pl_mass}. This figure shows that for all final protostellar masses $M_{\rm gap}$ is higher by a factor of ${\gtrsim}10$ in the early stages ($t\leq10^5$\,yr) than the late stages ($t\geq5 \times 10^5$\,yr). This holds for most radii above ${\sim}0.5$\,au. We also find that, at all times, $M_{\rm gap}$ decreases weakly (factor of ${\sim}2$) as the final protostellar mass increases, consistent with the findings of \cite{Sinclair2020} for Class II disks (where an age of 1 Myr was assumed).

Figure \ref{fig:min_pl_mass} also shows the typical radii from the star where super-Earth/mini-Neptune exoplanets are found (${\lesssim} 1$\,au) and those where the main population of giant exoplanets (ignoring hot Jupiters) is found (${\gtrsim} 1$\,au). This shows that even if a rocky planet formed completely in the Class 0/I stage ($t<2 \times 10^5$\,yr), it would either not produce an observable gap in the dusty disk or it would be just enough to start opening an observable gap. However, the same rocky planet could easily open a gap in a Class II disk. We note that the current observations of Class 0/I do not trace the inner 1\,au of the disk but those with the best angular resolution trace the disk on ${\sim}10$\,au scale (\citealt{Ohashi2023}). Future instruments such as the next generation Very Large Array, which operates at wavelengths where dust optical depth is less of a concern, may be better suited to tracing the impact of these smaller, closer-in, planets on their disks \citep{Bae2023}.

Thus moving to the more massive planets at larger radii (${\sim}10$\,au), Fig. \ref{fig:min_pl_mass} suggests that if a giant planet were completely formed in the early stages, it could open a gap in the embedded disks as well as the more mature disks. However, in the early stages, given that the gap and dusty rings are rarely observed for Class 0 disks (they are mostly observed for Class I disks and not all are due to planets; \citealt{Sheehan2020}; \citealt{Segura2020}; \citealt{Ohashi2023}; \citealt{Hsieh2024}), Fig. \ref{fig:min_pl_mass} puts an upper limit of ${\sim}50$\,M$_{\oplus}$ on the mass that can be accreted by planets in most disks during their Class 0 phase. Conversely, assuming that the substructures in some of the rare Class 0/I disks where gaps and dusty rings are observed are a result of planets, we would require a lower limit of ${\sim}10-50$\,M$_{\oplus}$ for the masses of those planets, i.e. ${\gtrsim}3-20\%$ of a Jupiter mass.
Since forming such massive protoplanets on these timescales will be challenging, in Sect. \ref{sec:implications} we estimate an upper limit on the fraction of disks that could produce such objects on these timescales.

\section{Discussion}
\label{sec:discuss}
\subsection{Implications for planet formation}
\label{sec:implications}
\begin{figure*}
    \centering
    \includegraphics[width=0.9\textwidth]{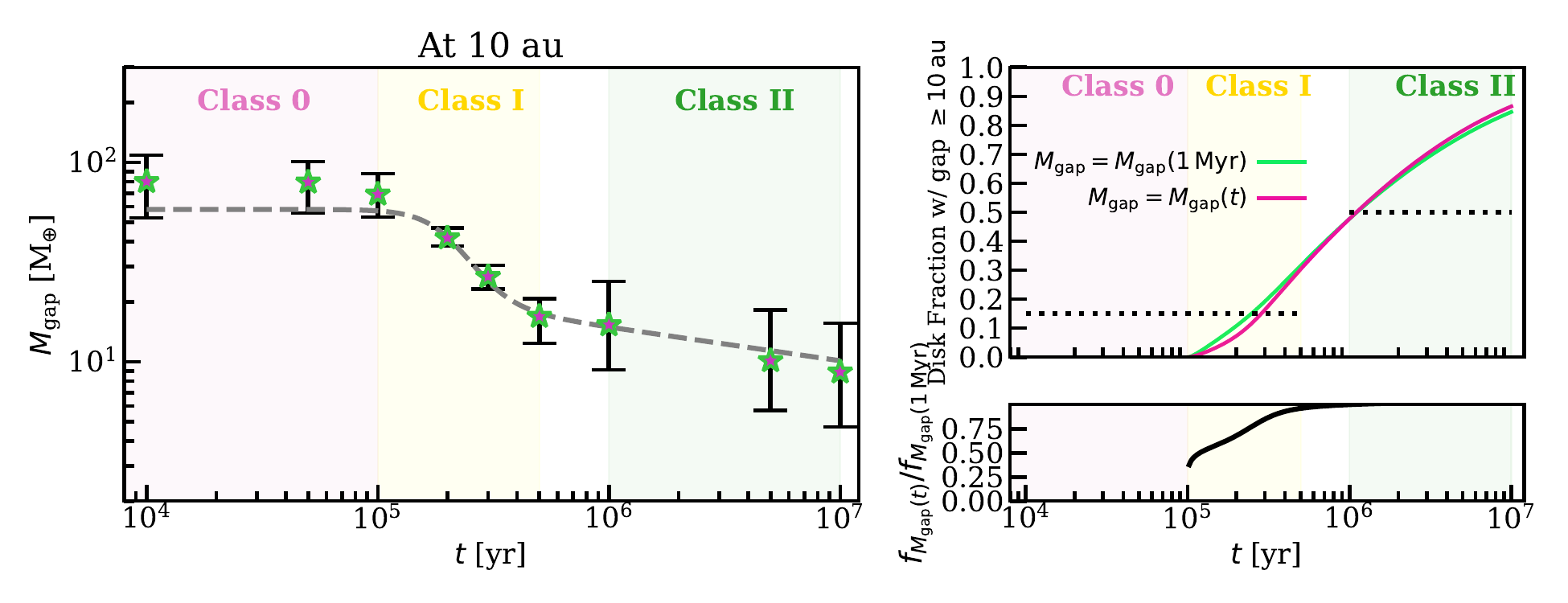}
    \caption{Left: The range of minimum gap-opening planet mass from the entire grid of models at 10\,au as a function of time. The stars show the mean at each time. A fitted function to the results of the models for $M_{\star, \rm f}=0.5$\,M$_{\odot}$ is shown in gray dashed line. Upper right: Fraction of disks that can open a gap at radii ${\geq}10$\,au if planets grow at the maximum rate allowed by pebble accretion (see text for details), assuming a gap-opening mass changing with time (solid pink line) found from the gray dashed line on the left panel. This provides an upper limit on the number of \textit{observable} substructures that are opened by planets. The green line shows the same but for a gap-opening planet mass at 1\,Myr (i.e., ${\sim}15$\,M$_{\oplus}$). The dotted black lines roughly indicate substructure occurrence rate in observed disks. Lower right: Pink line divided by the green line from the panel above showing the factor by which temperature effects alone reduce the number of gapped disks. Background shaded colors roughly show the different Classes.} 
    \label{fig:min_pl_mass_time}
\end{figure*}

Among the population of protoplanets inferred from gaps and cavities in disks assembled in the review by \citet{Bae2023}, their inferred semi-major axes are all ${\gtrsim}10\,\mathrm{au}$. Thus, the left panel of Fig.\,\ref{fig:min_pl_mass_time} summarizes how $M_{\rm gap}$ at $10\,\mathrm{au}$ changes over the course of a disk's evolution through the Class 0, I and II stages. The gap-opening mass stays relatively constant throughout Class 0, drops during Class I and finally declines very slowly over Class II. Motivated by this behavior, we fit an approximate functional form for $M_{\rm gap}(t)$, of which we provide further details in Appendix \ref{app:function}, using the $M_{\star, \rm f}=0.5$\,M$_{\odot}$ models. The range of $M_{\rm gap}$ amongst our Class II models is ${\sim}4-20$\,M$_{\oplus}$; the two lower/upper limits are consistent with the minimum estimated masses from the \citet{Bae2023} sample and from the updated estimates of \citet{Ruzza2024} respectively. Meanwhile, the largest gap-opening masses that we find are similar to the ${\sim}60$\,M$_{\oplus}$ median of the \citet{Bae2023} sample (while \citealt{Ruzza2024} generally estimate higher masses, with a median $\sim180\,M_{\oplus}$). Thus we might expect that, in the most extreme scenario where the entire population of putative protoplanets observed in Class II disks formed very rapidly at their inferred present locations, then the hotter temperatures and increased $M_{\rm gap}$ of younger embedded disks would effectively hide up to ${\sim}50\%$ of them, resulting in ${\gtrsim}25\%$ of embedded disks hosting substructures. Therefore, this alone is not enough to completely explain the difference in substructure frequency between embedded disks ($\lesssim10-15\%$) and Class II disks (${\gtrsim}50\%$). This suggests that either the substructures are roughly twice as hard to detect in embedded disks or the growth of planets must be ongoing between Class 0 and Class II disks.

Therefore, to estimate simply the \textit{relative} reduction in the fraction of growing planets which may open gaps in embedded disks due to their elevated temperatures, we use a toy model based on a modified version of the pebble accretion prescription of \citet[][Eq. 31]{Lambrechts2014b} (for full details see Appendix \ref{sec:pebble_accretion}; for discussion of other modes of planet formation see Appendix \ref{sec:grav_instabil}).
This is a model for 2D pebble accretion: though the core likely initially accretes in the 3D regime where its accretion radius is less than the pebble scale height, it should reach the 2D regime long before it can open a gap. Therefore we argue that the 2D regime is the relevant regime for determining the effects of delaying gap-opening until higher masses; the growth in the 3D regime will be independent of gap-opening mass. As our fiducial values, for all disks we assume an initial core mass $M_{\rm c,0}=10^{-1}\,M_{\oplus}$, to ensure the embryo accretes in the 2D regime \citep{Lambrechts2014b}, and formation time $t_{\rm i}=0.1\,\mathrm{Myr}$, such that the embryo is injected at the end of Class 0. This delay loosely accounts for various factors including the time for a) dust to settle out of the envelope/disk upper layers, b) conditions conducive to the streaming instability to arise and then the instability to grow - which can take hundreds of orbits \citep{Yang2017} even if the earlier stages are expedited by dust growth in the envelope (\citealt{Cridland2022}) and c) for the core to grow out of the 3D regime. We estimate that the embryos will spend less time in the 3D regime than the 2D regime, based on the relatively larger fraction of disks that can grow an embryo from $10^{-3}\,M_{\oplus}$ to $10^{-1}\,M_{\oplus}$ in the 3D regime than from $10^{-1}\,M_{\oplus}$ to $M_{\rm gap}(t)$ in the 2D regime (Appendix \ref{sec:pebble_accretion_3D}) so that should not be a limiting factor.
Moreover, there are alternative routes to forming cores which would start in the 2D regime via dust-assisted GI (\citealt{Longarini2023_sim, Longarini2023_theory, Baehr2023}; see Appendix \ref{sec:grav_instabil} for its caveats). Finally, we note that in this toy model planets are assumed to form in a smooth disk, without assistance from structures such as dust traps \citep[e.g.][]{Guilera2017,Morbidelli2020} or snowlines \citep[e.g.][]{Drazkowska2017}.

We use the pebble accretion prescription outlined above to determine the disk mass required to provide sufficient pebble flux to grow a planet of mass $M_{\rm gap}(t)$ in time $t$ (Eq. \ref{eq:S1_crit}-\ref{eq:Md_crit}), where $M_{\rm gap}(t)$ is given by our fitted function (Eq. \ref{eq:fitted}), and compare this to the case where $M_{\rm gap}$ is kept fixed at its value at 1 Myr.
We then use the lognormal fits to the Class 0 non-multiple dust mass distribution from the VANDAM survey \citep[][Tab. 9]{Tobin2020}, assuming a gas-to-dust ratio of $100$, in order to estimate how many disks started with enough mass to fulfill or exceed this pebble accretion criterion at each time in each case (Eq. \ref{eq:fgap_t}). We note that the masses presented by \cite{Tobin2020} are a conservative choice since the dust is likely to be optically thick at ALMA wavelengths (see \citealt{Tychoniec2018} who compare mass estimates from ALMA and VLA).

Our fiducial calculation is shown in the right panel of Fig. \ref{fig:min_pl_mass_time} (see Fig. \ref{fig:multi} for other versions assuming different input parameters). We carry out this analysis at 10 au, but this number can be interpreted as equivalent to the fraction of disks able to grow their furthest gap-opening planet at 10 au or greater - the range currently probed by observations e.g. \citep[][]{Ohashi2023,Bae2023} -  by time $t$. Planet growth is slower and gap opening is harder further from the star. However, disks that are massive enough to form a planet at 10\,au within time $t$ includes both disks that can form these planets \textit{exactly} in $t$ and disks that can form these planets in \textit{shorter} than $t$. Although the disks in the first category will not be able to produce gap-opening planets further than 10\,au in time $t$, the disks in the second category which are massive enough to grow gap-opening planets at 10\,au in \textit{shorter} than $t$, have the right mass to produce more massive gap-opening planets at some larger radius in \textit{exactly} $t$ (i.e. those disks susceptible to gap-opening further out by time $t$ are just a subset of those which can form gaps at 10 au).

The bottom-right panel of Fig.\,\ref{fig:min_pl_mass_time} shows that for the \citet{Lambrechts2014b} pebble fluxes, this calculation yields up to a factor of ${\sim}2-3$ decrease in the fraction of early Class I disks that have a high enough initial mass to produce an $M_{\rm gap}(t)$ planet given their elevated temperatures, compared to the same disks if they did not have elevated temperatures (i.e., using $M_{\rm gap}(1\,\mathrm{Myr})\approx15\,M_{\oplus}$ at all times).
For example, only ${\sim}5\%$ of disks start with enough mass to form a ${\gtrsim} 40\,M_{\oplus}$ planet in ${\lesssim}2\times10^5\,\mathrm{yr}$, compared to ${\sim}10\%$ with enough mass to form a ${\gtrsim} 10\,M_{\oplus}$ planet in that time. This suggests that ${\sim}50\%$ of planets which would otherwise be detectable in Class II would be hidden by the high disk temperatures in early Class I disks, consistent with that estimated above from the \citet{Bae2023} distribution of inferred protoplanet masses.

Moreover, by the middle of Class I (${\sim}0.3\,\mathrm{Myr}$) the predicted upper limit on disk fractions hosting gap-opening planets using the fiducial model (see top-right panel of Fig. \ref{fig:min_pl_mass_time}), is roughly in line with the observed frequency with which substructures are detected in embedded disks (${\lesssim}15\%$, \citealt{Ohashi2023, Hsieh2024}; where all might not be due to planets). We reiterate that our derived ${\sim}10-15\%$ value in top right panel of Fig. \ref{fig:min_pl_mass_time} is an upper limit on the number of \textit{observable} substructures due to planets since the opening of a gap in the gas surface density does not ensure its detectability due to opacity and geometrical effects. The fraction of disks that start with enough mass to have grown gap-opening planets continues to rise from 10-15\% in Class I into Class II, reaching ${\gtrsim}50\%$, consistent with observed substructure occurrence rates in Class II disks.

In Appendix \ref{sec:pebble_accretion}, we explore the sensitivity of these disk fractions to factors such as earlier embryo injection times, lower initial core masses, and reduced pebble fluxes\footnote{The pebble fluxes used in the prescription of \citet{Lambrechts2014b} are likely an upper limit since they are only valid in the limit where fragmentation does not limit grain sizes and so long as the pebbles do not grow beyond a Stokes number of 1 \citep{Drazkowska2021}.},
finding that the absolute values of disk fractions are model dependent. Considering those absolute values, in the most generous scenario that we explore, where massive cores form early and accrete from a high pebble flux (bottom right panel of Fig. \ref{fig:multi}), the toy model does not significantly exceed the fraction of observed disks with known substructures in early disks (especially when higher temperatures are considered) and thus current observations do not rule out such scenarios. Conversely, more conservative planet formation models that produce lower absolute fractions of substructured embedded disks (bottom left panel of Fig. \ref{fig:multi}) would suggest that a significant fraction of the observed substructures formed via mechanisms other than planets even in the Class II disks. We also note that based on our fiducial case (top-right panel of Fig. \ref{fig:min_pl_mass_time}) a high ($>80\%$) fraction of disks are expected to have substructures due to planets at 10\,Myr. Although this number is not in conflict with substructure detection rates of Class II disks, it may be perceived as inconsistent with current exoplanet population studies (e.g., \citealt{Rosenthal2022}). Ignoring the biases in exoplanet surveys (see Sect. \ref{sec:caveats}), this may be justified in our model if a time cut off of a few Myr was assumed. Moreover, we note that the \cite{Lambrechts2014b} prescription that we use, more strongly overestimates pebble fluxes at later times. This is because the pebble production line in their model moves outward in time to potentially infinity which is in conflict with the observed Class II disk radii. Therefore, a time cut off of a few Myr in our model is justifiable. In conclusion, the \textit{relative} decrease in the fraction of disks that can produce a gap-opening planet if elevated temperatures are considered (bottom right panel of Fig. \ref{fig:min_pl_mass_time}) seems to be a factor of ${\gtrsim}2$ regardless of the model assumptions. However, the \textit{absolute} fractions of disks themselves at a given time vary significantly depending on the assumptions made in the model (see Fig. \ref{fig:multi}).

Gas giant planets require runaway gas accretion to build their large gas masses \citep[e.g.][]{Pollack1996}. This phase happens after sufficient contraction of the gas envelope, a process which depends sensitively on the contribution of sublimating pebbles to the envelope's heating and opacity. Indeed, the few planets which could accrete ${\sim}10-50$\,M$_{\oplus}$ of their mass already in the protostellar stage (see Sect. \ref{sec:results}) suggest pebble accretion rates ${\sim}10^{-4}\,M_{\oplus}\,\mathrm{yr^{-1}}$, enough for pebbles to dominate the envelope opacity \citep{Brouwers2021}. Thus, runaway accretion likely requires planets to have reached their pebble isolation mass \citep[e.g.][]{Lambrechts2014a,Ormel2021} in order to ensure pebble accretion cannot prevent the cooling and contraction of the envelope.
Any giant planets that did form in embedded disks would therefore accrete a significant mass of pebbles, resulting in a high metallicity. However, structural modeling of gas giant mass-radius relationships implies metallicities $Z{\gtrsim}0.1$, equivalent to ${\sim}50-60\,M_{\oplus}$ of heavy elements for a $1\,M_{\rm J}$ planet \citep{Thorngren2016}; such a scenario is not therefore excluded by typical giant planet compositions.

\subsection{Caveats and comparative remarks}
\label{sec:caveats}

In this work we consider that the gas and dust are physically and thermally coupled. Moreover, we do not include dust settling and viscous heating. We also assume that planets can start forming in the protostellar phase. It is important to note that whether planets start forming early is actively under debate. Although some studies find that planets could start forming in the protostellar stage (\citealt{Tsukamoto2017,Tanaka2019, Cridland2022}; \citealt{Vorobyov2024}; \citealt{Baehr2023}; \citealt{Longarini2023_sim, Longarini2023_theory}), it is not yet clear whether they can without already existing substructures and perturbations (\citealt{Lyra2008,Guilera2017,Morbidelli2020,Voelkel2020,Lyra2024,sandor2024}). Further modeling work is needed to assess how this process could take place in the early stages which is beyond the scope of this letter.

We avoid making a direct comparison of the substructure occurrence rate in the Class 0/I phase with that of wide-orbit giant planets since much happens from that early phase until a planetary system is formed. For example, \cite{Lodato2019} (see also \citealt{Ndugu2019}) found that planet migration and evolution was necessary to produce a good match between the distribution of putative planets from gaps in Class II disks and the distribution of cold Jupiters. A meaningful comparison between current observations of Class 0/I disks and planets at radii ${\gtrsim}5$\,au is thus only possible with detailed modeling of the system's evolution, or a comparison of the trends of substructure type and planet mass (regardless of orbital radius) with stellar mass as done by \cite{vanderMarel2021} for mature disks, which are both beyond the scope of this letter. Moreover while disk observations do not yet resolve ${\lesssim}5$\,au where most Jupiter-mass planets have been discovered, it could also be argued that most of the current exoplanetary surveys are not yet sensitive enough to planets at larger radii of ${\gtrsim}5-10$\,au (e.g., see review by \citealt{Zhu2021}).   

Although the assumption for thermal coupling of dust and gas would break down in the disk surface layers (e.g., \citealt{Kamp2004}; \citealt{Bruderer2012}), it is a valid assumption in the disk mid-plane which has a high density and is the focus of this work. The assumption of no dust settling is likely realistic in the early stages given the observational evidence for small amount of settling in these disks (e.g., \citealt{Sheehan2022}; \citealt{vantHoff2023}; \citealt{Villenave2023}; \citealt{Encalada2024}). For mature disks this assumption is likely not valid and larger grains are expected to settle in the mid-plane (e.g., \citealt{Dullemond2004}; \citealt{Woitke2016}; \citealt{Villenave2020}). However, given the difference between the models with small grains and large grains in this work is small (Figs. \ref{fig:min_pl_mass} and \ref{fig:min_pl_mass_large}), including dust settling for mature disks will likely not change the conclusions and will not increase the $M_{\rm gap}$ by one order of magnitude.

Including viscous heating would increase the mid-plane temperature and thus increase the disk aspect ratio. However, if at all relevant, younger disks have larger accretion rates, thus are more likely to be influenced by viscous heating and higher mid-plane temperatures (e.g., \citealt{Manara2023}; \citealt{Fiorellino2023}; \citealt{Takakuwa2024}). Moreover, these young disks may be more turbulent \citep{Villenave2023} which could help close gaps and further reduce the number of substructures. Another effect of turbulence could be increased grain fragmentation (see \citealt{Testi2014} for a review), decreased grain sizes, and thus lower accretion rates onto the forming planets which would delay the gap formation further. Therefore, including viscous heating or turbulence for younger disks is expected to strengthen the conclusions made here. 

\section{Conclusions}

The difference in substructure frequency between embedded disks and older disks may imply unviability of early planet formation as an explanation for the insufficient inferred masses of Class II disks to form planetary systems. Motivated by this, we have considered the evolution of mid-plane temperature, disk aspect ratio, and the minimum planet mass needed to open a gap in the dust from early disks to mature ones. We found that Class 0/I disks have a higher mid-plane temperature and thus aspect ratio than Class II ones. We showed that a completely formed rocky planet at ${\sim}1$\,au or a giant planet with mass of up to ${\sim}50$\,M$_{\oplus}$ at 10\,au would not be able to open an observable gap in the Class 0/I stage, while that planet will easily open one in the Class II stage. We estimate that the number of planets hidden by elevated temperatures could be comparable to the number which are large enough to open gaps if they form early. Thus, the relative lack of substructures in young disks (where all the observed substructures might not be due to planets) may not imply such a low efficiency of early planet formation.

\begin{acknowledgements}
    We thank the referee for their constructive comments. We greatly appreciate the helpful discussions with C. Leedham and E.F. van Dishoeck. P.N. acknowledges support from the ESO Fellowship and IAU Gruber Foundation Fellowship programs.  
    A.D.S. was supported by funding from the European Research Council (ERC) under the European Union’s Horizon 2020 research and innovation programme (grant agreement No. 1010197S1 MOLDISK).
    G.P.R. is funded by the European Union under the European Union’s Horizon Europe Research \& Innovation Programme No. 101039651 (DiscEvol) and by the Fondazione Cariplo, grant no. 2022-1217. Views and opinions expressed are however those of the author(s) only and do not necessarily reflect those of the European Union or the European Research Council. Neither the European Union nor the granting authority can be held responsible for them.
\end{acknowledgements}

%
%

\bibliographystyle{aa}
\bibliography{gap_opening}

\begin{thebibliography}{150}
\expandafter\ifx\csname natexlab\endcsname\relax\def\natexlab#1{#1}\fi

\bibitem[{{Agurto-Gangas} {et~al.}(2019){Agurto-Gangas}, {Pineda}, {Sz{\H{u}}cs}, {Testi}, {Tazzari}, {Miotello}, {Caselli}, {Dunham}, {Stephens}, \& {Bourke}}]{Agurto2019}
{Agurto-Gangas}, C., {Pineda}, J.~E., {Sz{\H{u}}cs}, L., {et~al.} 2019, \aap, 623, A147

\bibitem[{{Akeson} {et~al.}(2013){Akeson}, {Chen}, {Ciardi}, {Crane}, {Good}, {Harbut}, {Jackson}, {Kane}, {Laity}, {Leifer}, {Lynn}, {McElroy}, {Papin}, {Plavchan}, {Ram{\'\i}rez}, {Rey}, {von Braun}, {Wittman}, {Abajian}, {Ali}, {Beichman}, {Beekley}, {Berriman}, {Berukoff}, {Bryden}, {Chan}, {Groom}, {Lau}, {Payne}, {Regelson}, {Saucedo}, {Schmitz}, {Stauffer}, {Wyatt}, \& {Zhang}}]{Akeson2013}
{Akeson}, R.~L., {Chen}, X., {Ciardi}, D., {et~al.} 2013, \pasp, 125, 989

\bibitem[{{Akiyama} {et~al.}(2016){Akiyama}, {Hasegawa}, {Hayashi}, \& {Iguchi}}]{Akiyama2016}
{Akiyama}, E., {Hasegawa}, Y., {Hayashi}, M., \& {Iguchi}, S. 2016, \apj, 818, 158

\bibitem[{{Alcal{\'a}} {et~al.}(2017){Alcal{\'a}}, {Manara}, {Natta}, {Frasca}, {Testi}, {Nisini}, {Stelzer}, {Williams}, {Antoniucci}, {Biazzo}, {Covino}, {Esposito}, {Getman}, \& {Rigliaco}}]{Alcala2017}
{Alcal{\'a}}, J.~M., {Manara}, C.~F., {Natta}, A., {et~al.} 2017, \aap, 600, A20

\bibitem[{{Anderson} {et~al.}(2024){Anderson}, {Cleeves}, {Blake}, {Qi}, {Bergin}, {Carpenter}, {Schwarz}, {Thilenius}, \& {Zhang}}]{Anderson2024}
{Anderson}, D.~E., {Cleeves}, L.~I., {Blake}, G.~A., {et~al.} 2024, \apj, 966, 84

\bibitem[{{Andrews} {et~al.}(2018){Andrews}, {Huang}, {P{\'e}rez}, {Isella}, {Dullemond}, {Kurtovic}, {Guzm{\'a}n}, {Carpenter}, {Wilner}, {Zhang}, {Zhu}, {Birnstiel}, {Bai}, {Benisty}, {Hughes}, {{\"O}berg}, \& {Ricci}}]{Andrews2018}
{Andrews}, S.~M., {Huang}, J., {P{\'e}rez}, L.~M., {et~al.} 2018, \apjl, 869, L41

\bibitem[{{Andrews} {et~al.}(2016){Andrews}, {Wilner}, {Zhu}, {Birnstiel}, {Carpenter}, {P{\'e}rez}, {Bai}, {{\"O}berg}, {Hughes}, {Isella}, \& {Ricci}}]{Andrews2016}
{Andrews}, S.~M., {Wilner}, D.~J., {Zhu}, Z., {et~al.} 2016, \apjl, 820, L40

\bibitem[{{Ansdell} {et~al.}(2016){Ansdell}, {Williams}, {van der Marel}, {Carpenter}, {Guidi}, {Hogerheijde}, {Mathews}, {Manara}, {Miotello}, {Natta}, {Oliveira}, {Tazzari}, {Testi}, {van Dishoeck}, \& {van Terwisga}}]{Ansdell2016}
{Ansdell}, M., {Williams}, J.~P., {van der Marel}, N., {et~al.} 2016, \apj, 828, 46

\bibitem[{{Armitage}(2010)}]{Armitage2010}
{Armitage}, P.~J. 2010, {Astrophysics of Planet Formation}

\bibitem[{{Ataiee} {et~al.}(2018){Ataiee}, {Baruteau}, {Alibert}, \& {Benz}}]{Ataiee2018}
{Ataiee}, S., {Baruteau}, C., {Alibert}, Y., \& {Benz}, W. 2018, \aap, 615, A110

\bibitem[{{Bae} {et~al.}(2023){Bae}, {Isella}, {Zhu}, {Martin}, {Okuzumi}, \& {Suriano}}]{Bae2023}
{Bae}, J., {Isella}, A., {Zhu}, Z., {et~al.} 2023, in Astronomical Society of the Pacific Conference Series, Vol. 534, Protostars and Planets VII, ed. S.~{Inutsuka}, Y.~{Aikawa}, T.~{Muto}, K.~{Tomida}, \& M.~{Tamura}, 423

\bibitem[{{Bae} {et~al.}(2017){Bae}, {Zhu}, \& {Hartmann}}]{Bae2017}
{Bae}, J., {Zhu}, Z., \& {Hartmann}, L. 2017, \apj, 850, 201

\bibitem[{{Baehr}(2023)}]{Baehr2023}
{Baehr}, H. 2023, \mnras, 523, 3348

\bibitem[{{Baraffe} {et~al.}(2002){Baraffe}, {Chabrier}, {Allard}, \& {Hauschildt}}]{Baraffe2002}
{Baraffe}, I., {Chabrier}, G., {Allard}, F., \& {Hauschildt}, P.~H. 2002, \aap, 382, 563

\bibitem[{{Baraffe} {et~al.}(2015){Baraffe}, {Homeier}, {Allard}, \& {Chabrier}}]{Baraffe2015}
{Baraffe}, I., {Homeier}, D., {Allard}, F., \& {Chabrier}, G. 2015, \aap, 577, A42

\bibitem[{{Baruteau} {et~al.}(2016){Baruteau}, {Bai}, {Mordasini}, \& {Molli{\`e}re}}]{Baruteau2016}
{Baruteau}, C., {Bai}, X., {Mordasini}, C., \& {Molli{\`e}re}, P. 2016, \ssr, 205, 77

\bibitem[{{Baruteau} {et~al.}(2014){Baruteau}, {Crida}, {Paardekooper}, {Masset}, {Guilet}, {Bitsch}, {Nelson}, {Kley}, \& {Papaloizou}}]{Baruteau2014}
{Baruteau}, C., {Crida}, A., {Paardekooper}, S.~J., {et~al.} 2014, in Protostars and Planets VI, ed. H.~{Beuther}, R.~S. {Klessen}, C.~P. {Dullemond}, \& T.~{Henning}, 667--689

\bibitem[{{Bate}(2022)}]{Bate2022}
{Bate}, M.~R. 2022, \mnras, 514, 2145

\bibitem[{{Belloche} {et~al.}(2020){Belloche}, {Maury}, {Maret}, {Anderl}, {Bacmann}, {Andr{\'e}}, {Bontemps}, {Cabrit}, {Codella}, {Gaudel}, {Gueth}, {Lef{\`e}vre}, {Lefloch}, {Podio}, \& {Testi}}]{Belloche2020}
{Belloche}, A., {Maury}, A.~J., {Maret}, S., {et~al.} 2020, \aap, 635, A198

\bibitem[{{Bitsch} {et~al.}(2015){Bitsch}, {Lambrechts}, \& {Johansen}}]{Bitsch2015b}
{Bitsch}, B., {Lambrechts}, M., \& {Johansen}, A. 2015, \aap, 582, A112

\bibitem[{{Bitsch} {et~al.}(2018){Bitsch}, {Morbidelli}, {Johansen}, {Lega}, {Lambrechts}, \& {Crida}}]{Bitsch2018}
{Bitsch}, B., {Morbidelli}, A., {Johansen}, A., {et~al.} 2018, \aap, 612, A30

\bibitem[{{Boss}(1997)}]{Boss1997}
{Boss}, A.~P. 1997, Science, 276, 1836

\bibitem[{{Brouwers} {et~al.}(2021){Brouwers}, {Ormel}, {Bonsor}, \& {Vazan}}]{Brouwers2021}
{Brouwers}, M.~G., {Ormel}, C.~W., {Bonsor}, A., \& {Vazan}, A. 2021, \aap, 653, A103

\bibitem[{{Bruderer} {et~al.}(2012){Bruderer}, {van Dishoeck}, {Doty}, \& {Herczeg}}]{Bruderer2012}
{Bruderer}, S., {van Dishoeck}, E.~F., {Doty}, S.~D., \& {Herczeg}, G.~J. 2012, \aap, 541, A91

\bibitem[{{Cacciapuoti} {et~al.}(2023){Cacciapuoti}, {Macias}, {Maury}, {Chandler}, {Sakai}, {Tychoniec}, {Viti}, {Natta}, {De Simone}, {Miotello}, {Codella}, {Ceccarelli}, {Podio}, {Fedele}, {Johnstone}, {Shirley}, {Liu}, {Bianchi}, {Zhang}, {Pineda}, {Loinard}, {M{\'e}nard}, {Lebreuilly}, {Klessen}, {Hennebelle}, {Molinari}, {Testi}, \& {Yamamoto}}]{Cacciapuoti2023}
{Cacciapuoti}, L., {Macias}, E., {Maury}, A.~J., {et~al.} 2023, \aap, 676, A4

\bibitem[{{Calvet} {et~al.}(2002){Calvet}, {D'Alessio}, {Hartmann}, {Wilner}, {Walsh}, \& {Sitko}}]{Calvet2002}
{Calvet}, N., {D'Alessio}, P., {Hartmann}, L., {et~al.} 2002, \apj, 568, 1008

\bibitem[{{Cieza} {et~al.}(2021){Cieza}, {Gonz{\'a}lez-Ruilova}, {Hales}, {Pinilla}, {Ru{\'\i}z-Rodr{\'\i}guez}, {Zurlo}, {Casassus}, {P{\'e}rez}, {C{\'a}novas}, {Arce-Tord}, {Flock}, {Kurtovic}, {Marino}, {Nogueira}, {Perez}, {Price}, {Principe}, \& {Williams}}]{Cieza2021}
{Cieza}, L.~A., {Gonz{\'a}lez-Ruilova}, C., {Hales}, A.~S., {et~al.} 2021, \mnras, 501, 2934

\bibitem[{{Crida} {et~al.}(2006){Crida}, {Morbidelli}, \& {Masset}}]{Crida2006}
{Crida}, A., {Morbidelli}, A., \& {Masset}, F. 2006, \icarus, 181, 587

\bibitem[{{Cridland} {et~al.}(2022){Cridland}, {Rosotti}, {Tabone}, {Tychoniec}, {McClure}, {Nazari}, \& {van Dishoeck}}]{Cridland2022}
{Cridland}, A.~J., {Rosotti}, G.~P., {Tabone}, B., {et~al.} 2022, \aap, 662, A90

\bibitem[{{Delussu} {et~al.}(2024){Delussu}, {Birnstiel}, {Miotello}, {Pinilla}, {Rosotti}, \& {Andrews}}]{Delussu2024}
{Delussu}, L., {Birnstiel}, T., {Miotello}, A., {et~al.} 2024, \aap, 688, A81

\bibitem[{{Dr{\k{a}}{\.z}kowska} \& {Alibert}(2017)}]{Drazkowska2017}
{Dr{\k{a}}{\.z}kowska}, J. \& {Alibert}, Y. 2017, \aap, 608, A92

\bibitem[{{Dr{\k{a}}{\.z}kowska} {et~al.}(2023){Dr{\k{a}}{\.z}kowska}, {Bitsch}, {Lambrechts}, {Mulders}, {Harsono}, {Vazan}, {Liu}, {Ormel}, {Kretke}, \& {Morbidelli}}]{Drkazkowska2023}
{Dr{\k{a}}{\.z}kowska}, J., {Bitsch}, B., {Lambrechts}, M., {et~al.} 2023, in Astronomical Society of the Pacific Conference Series, Vol. 534, Protostars and Planets VII, ed. S.~{Inutsuka}, Y.~{Aikawa}, T.~{Muto}, K.~{Tomida}, \& M.~{Tamura}, 717

\bibitem[{{Dr{\k{a}}{\.z}kowska} \& {Dullemond}(2018)}]{Drazkowska2018}
{Dr{\k{a}}{\.z}kowska}, J. \& {Dullemond}, C.~P. 2018, \aap, 614, A62

\bibitem[{{Dr{\k{a}}{\.z}kowska} {et~al.}(2021){Dr{\k{a}}{\.z}kowska}, {Stammler}, \& {Birnstiel}}]{Drazkowska2021}
{Dr{\k{a}}{\.z}kowska}, J., {Stammler}, S.~M., \& {Birnstiel}, T. 2021, \aap, 647, A15

\bibitem[{{Dullemond} \& {Dominik}(2004)}]{Dullemond2004}
{Dullemond}, C.~P. \& {Dominik}, C. 2004, \aap, 421, 1075

\bibitem[{{Dullemond} {et~al.}(2012){Dullemond}, {Juhasz}, {Pohl}, {Sereshti}, {Shetty}, {Peters}, {Commercon}, \& {Flock}}]{Dullemond2012}
{Dullemond}, C.~P., {Juhasz}, A., {Pohl}, A., {et~al.} 2012, {RADMC-3D: A multi-purpose radiative transfer tool}, Astrophysics Source Code Library, record ascl:1202.015

\bibitem[{{Dunham} {et~al.}(2010){Dunham}, {Evans}, {Terebey}, {Dullemond}, \& {Young}}]{Dunham2010}
{Dunham}, M.~M., {Evans}, Neal~J., I., {Terebey}, S., {Dullemond}, C.~P., \& {Young}, C.~H. 2010, \apj, 710, 470

\bibitem[{{Encalada} {et~al.}(2024){Encalada}, {Looney}, {Takakuwa}, {Tobin}, {Ohashi}, {J{\o}rgensen}, {Li}, {Aikawa}, {Aso}, {Koch}, {Kwon}, {Lai}, {Lee}, {Lin}, {Santamar{\'\i}a-Miranda}, {de Gregorio-Monsalvo}, {Phuong}, {Plunkett}, {Sai (Insa Choi)}, {Sharma}, {Yen}, \& {Han}}]{Encalada2024}
{Encalada}, F.~J., {Looney}, L.~W., {Takakuwa}, S., {et~al.} 2024, \apj, 966, 32

\bibitem[{{Evans} {et~al.}(2009){Evans}, {Dunham}, {J{\o}rgensen}, {Enoch}, {Mer{\'\i}n}, {van Dishoeck}, {Alcal{\'a}}, {Myers}, {Stapelfeldt}, {Huard}, {Allen}, {Harvey}, {van Kempen}, {Blake}, {Koerner}, {Mundy}, {Padgett}, \& {Sargent}}]{Evans2009}
{Evans}, Neal~J., I., {Dunham}, M.~M., {J{\o}rgensen}, J.~K., {et~al.} 2009, \apjs, 181, 321

\bibitem[{{Fedele} {et~al.}(2017){Fedele}, {Carney}, {Hogerheijde}, {Walsh}, {Miotello}, {Klaassen}, {Bruderer}, {Henning}, \& {van Dishoeck}}]{Fedele2017}
{Fedele}, D., {Carney}, M., {Hogerheijde}, M.~R., {et~al.} 2017, \aap, 600, A72

\bibitem[{{Fiorellino} {et~al.}(2023){Fiorellino}, {Tychoniec}, {Cruz-S{\'a}enz de Miera}, {Antoniucci}, {K{\'o}sp{\'a}l}, {Manara}, {Nisini}, \& {Rosotti}}]{Fiorellino2023}
{Fiorellino}, E., {Tychoniec}, {\L}., {Cruz-S{\'a}enz de Miera}, F., {et~al.} 2023, \apj, 944, 135

\bibitem[{{Fischer} {et~al.}(2017){Fischer}, {Megeath}, {Furlan}, {Ali}, {Stutz}, {Tobin}, {Osorio}, {Stanke}, {Manoj}, {Poteet}, {Booker}, {Hartmann}, {Wilson}, {Myers}, \& {Watson}}]{Fischer2017}
{Fischer}, W.~J., {Megeath}, S.~T., {Furlan}, E., {et~al.} 2017, \apj, 840, 69

\bibitem[{{Flock} {et~al.}(2015){Flock}, {Ruge}, {Dzyurkevich}, {Henning}, {Klahr}, \& {Wolf}}]{Flock2015}
{Flock}, M., {Ruge}, J.~P., {Dzyurkevich}, N., {et~al.} 2015, \aap, 574, A68

\bibitem[{{Francis} \& {van der Marel}(2020)}]{Francis2020}
{Francis}, L. \& {van der Marel}, N. 2020, \apj, 892, 111

\bibitem[{{Goldreich} {et~al.}(2004){Goldreich}, {Lithwick}, \& {Sari}}]{Goldreich2004}
{Goldreich}, P., {Lithwick}, Y., \& {Sari}, R. 2004, \araa, 42, 549

\bibitem[{{Guerra-Alvarado} {et~al.}(2024){Guerra-Alvarado}, {van der Marel}, {Di Francesco}, {Looney}, {Tobin}, {Cox}, {Sheehan}, {Wilner}, {Mac{\'\i}as}, \& {Carrasco-Gonz{\'a}lez}}]{Guerra-Alvarado2024}
{Guerra-Alvarado}, O.~M., {van der Marel}, N., {Di Francesco}, J., {et~al.} 2024, \aap, 681, A82

\bibitem[{{Guilera} \& {S{\'a}ndor}(2017)}]{Guilera2017}
{Guilera}, O.~M. \& {S{\'a}ndor}, Z. 2017, \aap, 604, A10

\bibitem[{{Harsono} {et~al.}(2018){Harsono}, {Bjerkeli}, {van der Wiel}, {Ramsey}, {Maud}, {Kristensen}, \& {J{\o}rgensen}}]{Harsono2018}
{Harsono}, D., {Bjerkeli}, P., {van der Wiel}, M. H.~D., {et~al.} 2018, Nature Astronomy, 2, 646

\bibitem[{{Harsono} {et~al.}(2015){Harsono}, {Bruderer}, \& {van Dishoeck}}]{Harsono2015}
{Harsono}, D., {Bruderer}, S., \& {van Dishoeck}, E.~F. 2015, \aap, 582, A41

\bibitem[{{Hartmann} {et~al.}(2016){Hartmann}, {Herczeg}, \& {Calvet}}]{Hartmann2016}
{Hartmann}, L., {Herczeg}, G., \& {Calvet}, N. 2016, \araa, 54, 135

\bibitem[{{Hsieh} {et~al.}(2024){Hsieh}, {Arce}, {Jos{\'e} Maureira}, {Pineda}, {Segura-Cox}, {Mardones}, {Dunham}, \& {Arun}}]{Hsieh2024}
{Hsieh}, C.-H., {Arce}, H.~G., {Jos{\'e} Maureira}, M., {et~al.} 2024, arXiv e-prints, arXiv:2404.02809

\bibitem[{{Johansen} {et~al.}(2014){Johansen}, {Blum}, {Tanaka}, {Ormel}, {Bizzarro}, \& {Rickman}}]{Johansen2014}
{Johansen}, A., {Blum}, J., {Tanaka}, H., {et~al.} 2014, in Protostars and Planets VI, ed. H.~{Beuther}, R.~S. {Klessen}, C.~P. {Dullemond}, \& T.~{Henning}, 547--570

\bibitem[{{Johansen} {et~al.}(2007){Johansen}, {Oishi}, {Mac Low}, {Klahr}, {Henning}, \& {Youdin}}]{Johansen2007}
{Johansen}, A., {Oishi}, J.~S., {Mac Low}, M.-M., {et~al.} 2007, \nat, 448, 1022

\bibitem[{{J{\o}rgensen} {et~al.}(2016){J{\o}rgensen}, {van der Wiel}, {Coutens}, {Lykke}, {M{\"u}ller}, {van Dishoeck}, {Calcutt}, {Bjerkeli}, {Bourke}, {Drozdovskaya}, {Favre}, {Fayolle}, {Garrod}, {Jacobsen}, {{\"O}berg}, {Persson}, \& {Wampfler}}]{Jorgensen2016}
{J{\o}rgensen}, J.~K., {van der Wiel}, M.~H.~D., {Coutens}, A., {et~al.} 2016, \aap, 595, A117

\bibitem[{{Kamp} \& {Dullemond}(2004)}]{Kamp2004}
{Kamp}, I. \& {Dullemond}, C.~P. 2004, \apj, 615, 991

\bibitem[{{Koga} {et~al.}(2022){Koga}, {Kawasaki}, \& {Machida}}]{Koga2022}
{Koga}, S., {Kawasaki}, Y., \& {Machida}, M.~N. 2022, \mnras, 515, 6073

\bibitem[{{Kratter} \& {Lodato}(2016)}]{Kratter2016}
{Kratter}, K. \& {Lodato}, G. 2016, \araa, 54, 271

\bibitem[{{Kratter} {et~al.}(2008){Kratter}, {Matzner}, \& {Krumholz}}]{Kratter2008}
{Kratter}, K.~M., {Matzner}, C.~D., \& {Krumholz}, M.~R. 2008, \apj, 681, 375

\bibitem[{{Kristensen} {et~al.}(2012){Kristensen}, {van Dishoeck}, {Bergin}, {Visser}, {Y{\i}ld{\i}z}, {San Jose-Garcia}, {J{\o}rgensen}, {Herczeg}, {Johnstone}, {Wampfler}, {Benz}, {Bruderer}, {Cabrit}, {Caselli}, {Doty}, {Harsono}, {Herpin}, {Hogerheijde}, {Karska}, {van Kempen}, {Liseau}, {Nisini}, {Tafalla}, {van der Tak}, \& {Wyrowski}}]{Kristensen2012}
{Kristensen}, L.~E., {van Dishoeck}, E.~F., {Bergin}, E.~A., {et~al.} 2012, \aap, 542, A8

\bibitem[{{Kwon} {et~al.}(2009){Kwon}, {Looney}, {Mundy}, {Chiang}, \& {Kemball}}]{Kwon2009}
{Kwon}, W., {Looney}, L.~W., {Mundy}, L.~G., {Chiang}, H.-F., \& {Kemball}, A.~J. 2009, \apj, 696, 841

\bibitem[{{Lambrechts} \& {Johansen}(2012)}]{Lambrechts2012}
{Lambrechts}, M. \& {Johansen}, A. 2012, \aap, 544, A32

\bibitem[{{Lambrechts} \& {Johansen}(2014)}]{Lambrechts2014b}
{Lambrechts}, M. \& {Johansen}, A. 2014, \aap, 572, A107

\bibitem[{{Lambrechts} {et~al.}(2014){Lambrechts}, {Johansen}, \& {Morbidelli}}]{Lambrechts2014a}
{Lambrechts}, M., {Johansen}, A., \& {Morbidelli}, A. 2014, \aap, 572, A35

\bibitem[{{Lau} {et~al.}(2024){Lau}, {Birnstiel}, {Dr{\k{a}}{\.z}kowska}, \& {Stammler}}]{Lau2024}
{Lau}, T. C.~H., {Birnstiel}, T., {Dr{\k{a}}{\.z}kowska}, J., \& {Stammler}, S. 2024, arXiv e-prints, arXiv:2406.12340

\bibitem[{{Law} {et~al.}(2022){Law}, {Crystian}, {Teague}, {{\"O}berg}, {Rich}, {Andrews}, {Bae}, {Flaherty}, {Guzm{\'a}n}, {Huang}, {Ilee}, {Kastner}, {Loomis}, {Long}, {P{\'e}rez}, {P{\'e}rez}, {Qi}, {Rosotti}, {Ru{\'\i}z-Rodr{\'\i}guez}, {Tsukagoshi}, \& {Wilner}}]{Law2022}
{Law}, C.~J., {Crystian}, S., {Teague}, R., {et~al.} 2022, \apj, 932, 114

\bibitem[{{Lebreuilly} {et~al.}(2023){Lebreuilly}, {Vallucci-Goy}, {Guillet}, {Lombart}, \& {Marchand}}]{Lebreuilly2023}
{Lebreuilly}, U., {Vallucci-Goy}, V., {Guillet}, V., {Lombart}, M., \& {Marchand}, P. 2023, \mnras, 518, 3326

\bibitem[{{Lee} {et~al.}(2022){Lee}, {Codella}, {Ceccarelli}, \& {L{\'o}pez-Sepulcre}}]{Lee2022}
{Lee}, C.-F., {Codella}, C., {Ceccarelli}, C., \& {L{\'o}pez-Sepulcre}, A. 2022, \apj, 937, 10

\bibitem[{{Lim} {et~al.}(2023){Lim}, {Simon}, {Li}, {Armitage}, {Carrera}, {Lyra}, {Rea}, {Yang}, \& {Youdin}}]{Lim2023}
{Lim}, J., {Simon}, J.~B., {Li}, R., {et~al.} 2023, arXiv e-prints, arXiv:2312.12508

\bibitem[{{Lin} \& {Papaloizou}(1993)}]{Lin1993}
{Lin}, D.~N.~C. \& {Papaloizou}, J.~C.~B. 1993, in Protostars and Planets III, ed. E.~H. {Levy} \& J.~I. {Lunine}, 749

\bibitem[{{Lin} {et~al.}(2023){Lin}, {Li}, {Tobin}, {Ohashi}, {J{\o}rgensen}, {Looney}, {Aso}, {Takakuwa}, {Aikawa}, {van't Hoff}, {de Gregorio-Monsalvo}, {Encalada}, {Flores}, {Gavino}, {Han}, {Kido}, {Koch}, {Kwon}, {Lai}, {Lee}, {Lee}, {Phuong}, {Sai}, {Sharma}, {Sheehan}, {Thieme}, {Williams}, {Yamato}, \& {Yen}}]{Lin2023}
{Lin}, Z.-Y.~D., {Li}, Z.-Y., {Tobin}, J.~J., {et~al.} 2023, \apj, 951, 9

\bibitem[{{Lissauer}(1993)}]{Lissauer1993}
{Lissauer}, J.~J. 1993, \araa, 31, 129

\bibitem[{{Liu} \& {Ji}(2020)}]{Liu2020}
{Liu}, B. \& {Ji}, J. 2020, Research in Astronomy and Astrophysics, 20, 164

\bibitem[{{Liu} {et~al.}(2019){Liu}, {Dipierro}, {Ragusa}, {Lodato}, {Herczeg}, {Long}, {Harsono}, {Boehler}, {Menard}, {Johnstone}, {Pascucci}, {Pinilla}, {Salyk}, {van der Plas}, {Cabrit}, {Fischer}, {Hendler}, {Manara}, {Nisini}, {Rigliaco}, {Avenhaus}, {Banzatti}, \& {Gully-Santiago}}]{Liu2019}
{Liu}, Y., {Dipierro}, G., {Ragusa}, E., {et~al.} 2019, \aap, 622, A75

\bibitem[{{Lodato} {et~al.}(2019){Lodato}, {Dipierro}, {Ragusa}, {Long}, {Herczeg}, {Pascucci}, {Pinilla}, {Manara}, {Tazzari}, {Liu}, {Mulders}, {Harsono}, {Boehler}, {M{\'e}nard}, {Johnstone}, {Salyk}, {van der Plas}, {Cabrit}, {Edwards}, {Fischer}, {Hendler}, {Nisini}, {Rigliaco}, {Avenhaus}, {Banzatti}, \& {Gully-Santiago}}]{Lodato2019}
{Lodato}, G., {Dipierro}, G., {Ragusa}, E., {et~al.} 2019, \mnras, 486, 453

\bibitem[{{Long} {et~al.}(2022){Long}, {Andrews}, {Rosotti}, {Harsono}, {Pinilla}, {Wilner}, {{\"O}berg}, {Teague}, {Trapman}, \& {Tabone}}]{Long2022}
{Long}, F., {Andrews}, S.~M., {Rosotti}, G., {et~al.} 2022, \apj, 931, 6

\bibitem[{{Long} {et~al.}(2018){Long}, {Pinilla}, {Herczeg}, {Harsono}, {Dipierro}, {Pascucci}, {Hendler}, {Tazzari}, {Ragusa}, {Salyk}, {Edwards}, {Lodato}, {van de Plas}, {Johnstone}, {Liu}, {Boehler}, {Cabrit}, {Manara}, {Menard}, {Mulders}, {Nisini}, {Fischer}, {Rigliaco}, {Banzatti}, {Avenhaus}, \& {Gully-Santiago}}]{Long2018}
{Long}, F., {Pinilla}, P., {Herczeg}, G.~J., {et~al.} 2018, \apj, 869, 17

\bibitem[{{Long} {et~al.}(2023){Long}, {Ren}, {Wallack}, {Harsono}, {Herczeg}, {Pinilla}, {Mawet}, {Liu}, {Andrews}, {Bai}, {Cabrit}, {Cieza}, {Johnstone}, {Leisenring}, {Lodato}, {Liu}, {Manara}, {Mulders}, {Ragusa}, {Sallum}, {Shi}, {Tazzari}, {Uyama}, {Wagner}, {Wilner}, \& {Xuan}}]{Long2023}
{Long}, F., {Ren}, B.~B., {Wallack}, N.~L., {et~al.} 2023, \apj, 949, 27

\bibitem[{{Longarini} {et~al.}(2023{\natexlab{a}}){Longarini}, {Armitage}, {Lodato}, {Price}, \& {Ceppi}}]{Longarini2023_sim}
{Longarini}, C., {Armitage}, P.~J., {Lodato}, G., {Price}, D.~J., \& {Ceppi}, S. 2023{\natexlab{a}}, \mnras, 522, 6217

\bibitem[{{Longarini} {et~al.}(2023{\natexlab{b}}){Longarini}, {Lodato}, {Bertin}, \& {Armitage}}]{Longarini2023_theory}
{Longarini}, C., {Lodato}, G., {Bertin}, G., \& {Armitage}, P.~J. 2023{\natexlab{b}}, \mnras, 519, 2017

\bibitem[{{Loomis} {et~al.}(2020){Loomis}, {{\"O}berg}, {Andrews}, {Bergin}, {Bergner}, {Blake}, {Cleeves}, {Czekala}, {Huang}, {Le Gal}, {M{\'e}nard}, {Pegues}, {Qi}, {Walsh}, {Williams}, \& {Wilner}}]{Loomis2020}
{Loomis}, R.~A., {{\"O}berg}, K.~I., {Andrews}, S.~M., {et~al.} 2020, \apj, 893, 101

\bibitem[{{Lyra} {et~al.}(2008){Lyra}, {Johansen}, {Klahr}, \& {Piskunov}}]{Lyra2008}
{Lyra}, W., {Johansen}, A., {Klahr}, H., \& {Piskunov}, N. 2008, \aap, 491, L41

\bibitem[{{Lyra} {et~al.}(2024){Lyra}, {Yang}, {Simon}, {Umurhan}, \& {Youdin}}]{Lyra2024}
{Lyra}, W., {Yang}, C.-C., {Simon}, J.~B., {Umurhan}, O.~M., \& {Youdin}, A.~N. 2024, arXiv e-prints, arXiv:2406.17934

\bibitem[{{Manara} {et~al.}(2023){Manara}, {Ansdell}, {Rosotti}, {Hughes}, {Armitage}, {Lodato}, \& {Williams}}]{Manara2023}
{Manara}, C.~F., {Ansdell}, M., {Rosotti}, G.~P., {et~al.} 2023, in Astronomical Society of the Pacific Conference Series, Vol. 534, Protostars and Planets VII, ed. S.~{Inutsuka}, Y.~{Aikawa}, T.~{Muto}, K.~{Tomida}, \& M.~{Tamura}, 539

\bibitem[{{Manara} {et~al.}(2018){Manara}, {Morbidelli}, \& {Guillot}}]{Manara2018}
{Manara}, C.~F., {Morbidelli}, A., \& {Guillot}, T. 2018, \aap, 618, L3

\bibitem[{{Meru} {et~al.}(2019){Meru}, {Rosotti}, {Booth}, {Nazari}, \& {Clarke}}]{Meru2019}
{Meru}, F., {Rosotti}, G.~P., {Booth}, R.~A., {Nazari}, P., \& {Clarke}, C.~J. 2019, \mnras, 482, 3678

\bibitem[{{Meyer} {et~al.}(1997){Meyer}, {Calvet}, \& {Hillenbrand}}]{Meyer1997}
{Meyer}, M.~R., {Calvet}, N., \& {Hillenbrand}, L.~A. 1997, \aj, 114, 288

\bibitem[{{Miotello} {et~al.}(2016){Miotello}, {van Dishoeck}, {Kama}, \& {Bruderer}}]{Miotello2016}
{Miotello}, A., {van Dishoeck}, E.~F., {Kama}, M., \& {Bruderer}, S. 2016, \aap, 594, A85

\bibitem[{{Morbidelli}(2020)}]{Morbidelli2020}
{Morbidelli}, A. 2020, \aap, 638, A1

\bibitem[{{Najita} \& {Kenyon}(2014)}]{Najita2014}
{Najita}, J.~R. \& {Kenyon}, S.~J. 2014, \mnras, 445, 3315

\bibitem[{{Narang} {et~al.}(2024){Narang}, {Manoj}, {Tyagi}, {Watson}, {Megeath}, {Federman}, {Rubinstein}, {Gutermuth}, {Caratti o Garatti}, {Beuther}, {Bourke}, {Van Dishoeck}, {Evans}, {Anglada}, {Osorio}, {Stanke}, {Muzerolle}, {Looney}, {Yang}, {Klaassen}, {Karnath}, {Atnagulov}, {Brunken}, {Fischer}, {Furlan}, {Green}, {Habel}, {Hartmann}, {Linz}, {Nazari}, {Pokhrel}, {Rahatgaonkar}, {Rocha}, {Sheehan}, {Slavicinska}, {Stutz}, {Tobin}, {Tychoniec}, \& {Wolk}}]{Narang2024}
{Narang}, M., {Manoj}, P., {Tyagi}, H., {et~al.} 2024, \apjl, 962, L16

\bibitem[{{Nazari} {et~al.}(2024{\natexlab{a}}){Nazari}, {Cheung}, {Ferrer Asensio}, {Murillo}, {van Dishoeck}, {J{\o}rgensen}, {Bourke}, {Chuang}, {Drozdovskaya}, {Fedoseev}, {Garrod}, {Ioppolo}, {Linnartz}, {McGuire}, {M{\"u}ller}, {Qasim}, \& {Wampfler}}]{Nazari2024}
{Nazari}, P., {Cheung}, J.~S.~Y., {Ferrer Asensio}, J., {et~al.} 2024{\natexlab{a}}, arXiv e-prints, arXiv:2401.04760

\bibitem[{{Nazari} {et~al.}(2024{\natexlab{b}}){Nazari}, {Tabone}, {Ahmadi}, {Cabrit}, {van Dishoeck}, {Codella}, {Ferreira}, {Podio}, {Tychoniec}, \& {van Gelder}}]{Nazari2024_wind}
{Nazari}, P., {Tabone}, B., {Ahmadi}, A., {et~al.} 2024{\natexlab{b}}, arXiv e-prints, arXiv:2402.18631

\bibitem[{{Nazari} {et~al.}(2023){Nazari}, {Tabone}, \& {Rosotti}}]{Nazari2023}
{Nazari}, P., {Tabone}, B., \& {Rosotti}, G.~P. 2023, \aap, 671, A107

\bibitem[{{Nazari} {et~al.}(2022){Nazari}, {Tabone}, {Rosotti}, {van Gelder}, {Meshaka}, \& {van Dishoeck}}]{Nazari2022_disk}
{Nazari}, P., {Tabone}, B., {Rosotti}, G.~P., {et~al.} 2022, \aap, 663, A58

\bibitem[{{Ndugu} {et~al.}(2019){Ndugu}, {Bitsch}, \& {Jurua}}]{Ndugu2019}
{Ndugu}, N., {Bitsch}, B., \& {Jurua}, E. 2019, \mnras, 488, 3625

\bibitem[{{Ohashi} {et~al.}(2023){Ohashi}, {Tobin}, {J{\o}rgensen}, {Takakuwa}, {Sheehan}, {Aikawa}, {Li}, {Looney}, {Williams}, {Aso}, {Sharma}, {Sai}, {Yamato}, {Lee}, {Tomida}, {Yen}, {Encalada}, {Flores}, {Gavino}, {Kido}, {Han}, {Lin}, {Narayanan}, {Phuong}, {Santamar{\'\i}a-Miranda}, {Thieme}, {van't Hoff}, {de Gregorio-Monsalvo}, {Koch}, {Kwon}, {Lai}, {Lee}, {Plunkett}, {Saigo}, {Hirano}, {Lam}, \& {Mori}}]{Ohashi2023}
{Ohashi}, N., {Tobin}, J.~J., {J{\o}rgensen}, J.~K., {et~al.} 2023, \apj, 951, 8

\bibitem[{{Ormel} \& {Klahr}(2010)}]{Ormel2010}
{Ormel}, C.~W. \& {Klahr}, H.~H. 2010, \aap, 520, A43

\bibitem[{{Ormel} {et~al.}(2021){Ormel}, {Vazan}, \& {Brouwers}}]{Ormel2021}
{Ormel}, C.~W., {Vazan}, A., \& {Brouwers}, M.~G. 2021, \aap, 647, A175

\bibitem[{{Ossenkopf} \& {Henning}(1994)}]{Ossenkopf1994}
{Ossenkopf}, V. \& {Henning}, T. 1994, \aap, 291, 943

\bibitem[{{Paardekooper} \& {Mellema}(2006)}]{Paardekooper2006}
{Paardekooper}, S.~J. \& {Mellema}, G. 2006, \aap, 459, L17

\bibitem[{{Paneque-Carre{\~n}o} {et~al.}(2022){Paneque-Carre{\~n}o}, {Miotello}, {van Dishoeck}, {P{\'e}rez}, {Facchini}, {Izquierdo}, {Tychoniec}, \& {Testi}}]{Paneque2022}
{Paneque-Carre{\~n}o}, T., {Miotello}, A., {van Dishoeck}, E.~F., {et~al.} 2022, \aap, 666, A168

\bibitem[{{Pinilla} {et~al.}(2018){Pinilla}, {Tazzari}, {Pascucci}, {Youdin}, {Garufi}, {Manara}, {Testi}, {van der Plas}, {Barenfeld}, {Canovas}, {Cox}, {Hendler}, {P{\'e}rez}, \& {van der Marel}}]{Pinilla2018}
{Pinilla}, P., {Tazzari}, M., {Pascucci}, I., {et~al.} 2018, \apj, 859, 32

\bibitem[{{Pinte} {et~al.}(2018){Pinte}, {M{\'e}nard}, {Duch{\^e}ne}, {Hill}, {Dent}, {Woitke}, {Maret}, {van der Plas}, {Hales}, {Kamp}, {Thi}, {de Gregorio-Monsalvo}, {Rab}, {Quanz}, {Avenhaus}, {Carmona}, \& {Casassus}}]{Pinte2018}
{Pinte}, C., {M{\'e}nard}, F., {Duch{\^e}ne}, G., {et~al.} 2018, \aap, 609, A47

\bibitem[{{Pokhrel} {et~al.}(2023){Pokhrel}, {Megeath}, {Gutermuth}, {Furlan}, {Fischer}, {Federman}, {Tobin}, {Stutz}, {Hartmann}, {Osorio}, {Watson}, {Stanke}, {Manoj}, {Narang}, {Atnagulov}, {Habel}, \& {Zakri}}]{Pokhrel2023}
{Pokhrel}, R., {Megeath}, S.~T., {Gutermuth}, R.~A., {et~al.} 2023, \apjs, 266, 32

\bibitem[{{Pollack} {et~al.}(1996){Pollack}, {Hubickyj}, {Bodenheimer}, {Lissauer}, {Podolak}, \& {Greenzweig}}]{Pollack1996}
{Pollack}, J.~B., {Hubickyj}, O., {Bodenheimer}, P., {et~al.} 1996, \icarus, 124, 62

\bibitem[{{Qiao} {et~al.}(2023){Qiao}, {Coleman}, \& {Haworth}}]{Qiao2023}
{Qiao}, L., {Coleman}, G. A.~L., \& {Haworth}, T.~J. 2023, \mnras, 522, 1939

\bibitem[{{Rice} {et~al.}(2011){Rice}, {Armitage}, {Mamatsashvili}, {Lodato}, \& {Clarke}}]{Rice2011}
{Rice}, W.~K.~M., {Armitage}, P.~J., {Mamatsashvili}, G.~R., {Lodato}, G., \& {Clarke}, C.~J. 2011, \mnras, 418, 1356

\bibitem[{{Rosenthal} {et~al.}(2022){Rosenthal}, {Knutson}, {Chachan}, {Dai}, {Howard}, {Fulton}, {Chontos}, {Crepp}, {Dalba}, {Henry}, {Kane}, {Petigura}, {Weiss}, \& {Wright}}]{Rosenthal2022}
{Rosenthal}, L.~J., {Knutson}, H.~A., {Chachan}, Y., {et~al.} 2022, \apjs, 262, 1

\bibitem[{{Rosotti} {et~al.}(2016){Rosotti}, {Juhasz}, {Booth}, \& {Clarke}}]{Rosotti2016}
{Rosotti}, G.~P., {Juhasz}, A., {Booth}, R.~A., \& {Clarke}, C.~J. 2016, \mnras, 459, 2790

\bibitem[{{Ruzza} {et~al.}(2024){Ruzza}, {Lodato}, \& {Rosotti}}]{Ruzza2024}
{Ruzza}, A., {Lodato}, G., \& {Rosotti}, G.~P. 2024, \aap, 685, A65

\bibitem[{{S{\'a}ndor} {et~al.}(2024){S{\'a}ndor}, {Guilera}, {Reg{\'a}ly}, \& {Lyra}}]{sandor2024}
{S{\'a}ndor}, Z., {Guilera}, O.~M., {Reg{\'a}ly}, Z., \& {Lyra}, W. 2024, \aap, 686, A78

\bibitem[{{Savvidou} \& {Bitsch}(2023)}]{Savvidou2023}
{Savvidou}, S. \& {Bitsch}, B. 2023, \aap, 679, A42

\bibitem[{{Savvidou} \& {Bitsch}(2024)}]{Savvidou2024}
{Savvidou}, S. \& {Bitsch}, B. 2024, arXiv e-prints, arXiv:2407.08533

\bibitem[{{Segura-Cox} {et~al.}(2020){Segura-Cox}, {Schmiedeke}, {Pineda}, {Stephens}, {Fern{\'a}ndez-L{\'o}pez}, {Looney}, {Caselli}, {Li}, {Mundy}, {Kwon}, \& {Harris}}]{Segura2020}
{Segura-Cox}, D.~M., {Schmiedeke}, A., {Pineda}, J.~E., {et~al.} 2020, \nat, 586, 228

\bibitem[{{Sellek} {et~al.}(2020){Sellek}, {Booth}, \& {Clarke}}]{Sellek2020}
{Sellek}, A.~D., {Booth}, R.~A., \& {Clarke}, C.~J. 2020, \mnras, 498, 2845

\bibitem[{{Shakura} \& {Sunyaev}(1973)}]{Shakura1973}
{Shakura}, N.~I. \& {Sunyaev}, R.~A. 1973, \aap, 24, 337

\bibitem[{{Sharma} {et~al.}(2023){Sharma}, {J{\o}rgensen}, {Gavino}, {Ohashi}, {Tobin}, {Lin}, {Li}, {Takakuwa}, {Lee}, {Sai (Insa Choi)}, {Kwon}, {de Gregorio-Monsalvo}, {Santamar{\'\i}a-Miranda}, {Yen}, {Aikawa}, {Aso}, {Lai}, {Lee}, {Looney}, {Phuong}, {Thieme}, \& {Williams}}]{Sharma2023}
{Sharma}, R., {J{\o}rgensen}, J.~K., {Gavino}, S., {et~al.} 2023, \apj, 954, 69

\bibitem[{{Sheehan} \& {Eisner}(2018)}]{Sheehan2018}
{Sheehan}, P.~D. \& {Eisner}, J.~A. 2018, \apj, 857, 18

\bibitem[{{Sheehan} {et~al.}(2020){Sheehan}, {Tobin}, {Federman}, {Megeath}, \& {Looney}}]{Sheehan2020}
{Sheehan}, P.~D., {Tobin}, J.~J., {Federman}, S., {Megeath}, S.~T., \& {Looney}, L.~W. 2020, \apj, 902, 141

\bibitem[{{Sheehan} {et~al.}(2022){Sheehan}, {Tobin}, {Li}, {van't Hoff}, {J{\o}rgensen}, {Kwon}, {Looney}, {Ohashi}, {Takakuwa}, {Williams}, {Aso}, {Gavino}, {de Gregorio-Monsalvo}, {Han}, {Lee}, {Plunkett}, {Sharma}, {Aikawa}, {Lai}, {Lee}, {Lin}, {Saigo}, {Tomida}, \& {Yen}}]{Sheehan2022}
{Sheehan}, P.~D., {Tobin}, J.~J., {Li}, Z.-Y., {et~al.} 2022, \apj, 934, 95

\bibitem[{{Sinclair} {et~al.}(2020){Sinclair}, {Rosotti}, {Juhasz}, \& {Clarke}}]{Sinclair2020}
{Sinclair}, C.~A., {Rosotti}, G.~P., {Juhasz}, A., \& {Clarke}, C.~J. 2020, \mnras, 493, 3535

\bibitem[{{Takahashi} \& {Inutsuka}(2016)}]{Takahashi2016}
{Takahashi}, S.~Z. \& {Inutsuka}, S.-i. 2016, \aj, 152, 184

\bibitem[{{Takakuwa} {et~al.}(2024){Takakuwa}, {Saigo}, {Kido}, {Ohashi}, {Tobin}, {J{\o}rgensen}, {Aikawa}, {Aso}, {Gavino}, {Han}, {Koch}, {Kwon}, {Lee}, {Lee}, {Li}, {Lin}, {Looney}, {Mori}, {Sai}, {Sharma}, {Sheehan}, {Tomida}, {Williams}, {Yamato}, \& {Yen}}]{Takakuwa2024}
{Takakuwa}, S., {Saigo}, K., {Kido}, M., {et~al.} 2024, \apj, 964, 24

\bibitem[{{Tanaka} \& {Tsukamoto}(2019)}]{Tanaka2019}
{Tanaka}, Y.~A. \& {Tsukamoto}, Y. 2019, \mnras, 484, 1574

\bibitem[{{Testi} {et~al.}(2014){Testi}, {Birnstiel}, {Ricci}, {Andrews}, {Blum}, {Carpenter}, {Dominik}, {Isella}, {Natta}, {Williams}, \& {Wilner}}]{Testi2014}
{Testi}, L., {Birnstiel}, T., {Ricci}, L., {et~al.} 2014, in Protostars and Planets VI, ed. H.~{Beuther}, R.~S. {Klessen}, C.~P. {Dullemond}, \& T.~{Henning}, 339--361

\bibitem[{{Thi} {et~al.}(2004){Thi}, {van Zadelhoff}, \& {van Dishoeck}}]{Thi2004}
{Thi}, W.~F., {van Zadelhoff}, G.~J., \& {van Dishoeck}, E.~F. 2004, \aap, 425, 955

\bibitem[{{Thorngren} {et~al.}(2016){Thorngren}, {Fortney}, {Murray-Clay}, \& {Lopez}}]{Thorngren2016}
{Thorngren}, D.~P., {Fortney}, J.~J., {Murray-Clay}, R.~A., \& {Lopez}, E.~D. 2016, \apj, 831, 64

\bibitem[{{Tobin} {et~al.}(2022){Tobin}, {Offner}, {Kratter}, {Megeath}, {Sheehan}, {Looney}, {Diaz-Rodriguez}, {Osorio}, {Anglada}, {Sadavoy}, {Furlan}, {Segura-Cox}, {Karnath}, {van't Hoff}, {van Dishoeck}, {Li}, {Sharma}, {Stutz}, \& {Tychoniec}}]{Tobin2022}
{Tobin}, J.~J., {Offner}, S. S.~R., {Kratter}, K.~M., {et~al.} 2022, \apj, 925, 39

\bibitem[{{Tobin} \& {Sheehan}(2024)}]{Tobin2024}
{Tobin}, J.~J. \& {Sheehan}, P.~D. 2024, arXiv e-prints, arXiv:2403.15550

\bibitem[{{Tobin} {et~al.}(2020){Tobin}, {Sheehan}, {Megeath}, {D{\'\i}az-Rodr{\'\i}guez}, {Offner}, {Murillo}, {van 't Hoff}, {van Dishoeck}, {Osorio}, {Anglada}, {Furlan}, {Stutz}, {Reynolds}, {Karnath}, {Fischer}, {Persson}, {Looney}, {Li}, {Stephens}, {Chandler}, {Cox}, {Dunham}, {Tychoniec}, {Kama}, {Kratter}, {Kounkel}, {Mazur}, {Maud}, {Patel}, {Perez}, {Sadavoy}, {Segura-Cox}, {Sharma}, {Stephenson}, {Watson}, \& {Wyrowski}}]{Tobin2020}
{Tobin}, J.~J., {Sheehan}, P.~D., {Megeath}, S.~T., {et~al.} 2020, \apj, 890, 130

\bibitem[{{Tsukamoto} {et~al.}(2017){Tsukamoto}, {Okuzumi}, \& {Kataoka}}]{Tsukamoto2017}
{Tsukamoto}, Y., {Okuzumi}, S., \& {Kataoka}, A. 2017, \apj, 838, 151

\bibitem[{{Tychoniec} {et~al.}(2020){Tychoniec}, {Manara}, {Rosotti}, {van Dishoeck}, {Cridland}, {Hsieh}, {Murillo}, {Segura-Cox}, {van Terwisga}, \& {Tobin}}]{Tychoniec2020}
{Tychoniec}, {\L}., {Manara}, C.~F., {Rosotti}, G.~P., {et~al.} 2020, \aap, 640, A19

\bibitem[{{Tychoniec} {et~al.}(2018){Tychoniec}, {Tobin}, {Karska}, {Chandler}, {Dunham}, {Harris}, {Kratter}, {Li}, {Looney}, {Melis}, {P{\'e}rez}, {Sadavoy}, {Segura-Cox}, \& {van Dishoeck}}]{Tychoniec2018}
{Tychoniec}, {\L}., {Tobin}, J.~J., {Karska}, A., {et~al.} 2018, \apjs, 238, 19

\bibitem[{{Tzouvanou} {et~al.}(2023){Tzouvanou}, {Bitsch}, \& {Pichierri}}]{Tzouvanou2023}
{Tzouvanou}, A., {Bitsch}, B., \& {Pichierri}, G. 2023, \aap, 677, A82

\bibitem[{{van der Marel} \& {Mulders}(2021)}]{vanderMarel2021}
{van der Marel}, N. \& {Mulders}, G.~D. 2021, \aj, 162, 28

\bibitem[{{van't Hoff} {et~al.}(2020){van't Hoff}, {Harsono}, {Tobin}, {Bosman}, {van Dishoeck}, {J{\o}rgensen}, {Miotello}, {Murillo}, \& {Walsh}}]{vantHoff2020}
{van't Hoff}, M. L.~R., {Harsono}, D., {Tobin}, J.~J., {et~al.} 2020, \apj, 901, 166

\bibitem[{{van't Hoff} {et~al.}(2023){van't Hoff}, {Tobin}, {Li}, {Ohashi}, {J{\o}rgensen}, {Lin}, {Aikawa}, {Aso}, {de Gregorio-Monsalvo}, {Gavino}, {Han}, {Koch}, {Kwon}, {Lee}, {Lee}, {Looney}, {Narayanan}, {Plunkett}, {Sai}, {Santamar{\'\i}a-Miranda}, {Sharma}, {Sheehan}, {Takakuwa}, {Thieme}, {Williams}, {Lai}, {Phuong}, \& {Yen}}]{vantHoff2023}
{van't Hoff}, M. L.~R., {Tobin}, J.~J., {Li}, Z.-Y., {et~al.} 2023, \apj, 951, 10

\bibitem[{{Villenave} {et~al.}(2020){Villenave}, {M{\'e}nard}, {Dent}, {Duch{\^e}ne}, {Stapelfeldt}, {Benisty}, {Boehler}, {van der Plas}, {Pinte}, {Telkamp}, {Wolff}, {Flores}, {Lesur}, {Louvet}, {Riols}, {Dougados}, {Williams}, \& {Padgett}}]{Villenave2020}
{Villenave}, M., {M{\'e}nard}, F., {Dent}, W.~R.~F., {et~al.} 2020, \aap, 642, A164

\bibitem[{{Villenave} {et~al.}(2023){Villenave}, {Podio}, {Duch{\^e}ne}, {Stapelfeldt}, {Melis}, {Carrasco-Gonzalez}, {Le Gouellec}, {M{\'e}nard}, {de Simone}, {Chandler}, {Garufi}, {Pinte}, {Bianchi}, \& {Codella}}]{Villenave2023}
{Villenave}, M., {Podio}, L., {Duch{\^e}ne}, G., {et~al.} 2023, \apj, 946, 70

\bibitem[{{Visser} {et~al.}(2009){Visser}, {van Dishoeck}, {Doty}, \& {Dullemond}}]{Visser2009}
{Visser}, R., {van Dishoeck}, E.~F., {Doty}, S.~D., \& {Dullemond}, C.~P. 2009, \aap, 495, 881

\bibitem[{{Voelkel} {et~al.}(2020){Voelkel}, {Klahr}, {Mordasini}, {Emsenhuber}, \& {Lenz}}]{Voelkel2020}
{Voelkel}, O., {Klahr}, H., {Mordasini}, C., {Emsenhuber}, A., \& {Lenz}, C. 2020, \aap, 642, A75

\bibitem[{{Vorobyov} {et~al.}(2024){Vorobyov}, {Kulikov}, {Elbakyan}, {McKevitt}, \& {G{\"u}del}}]{Vorobyov2024}
{Vorobyov}, E.~I., {Kulikov}, I., {Elbakyan}, V.~G., {McKevitt}, J., \& {G{\"u}del}, M. 2024, \aap, 683, A202

\bibitem[{{Winter} \& {Haworth}(2022)}]{Winter2022}
{Winter}, A.~J. \& {Haworth}, T.~J. 2022, European Physical Journal Plus, 137, 1132

\bibitem[{{Woitke} {et~al.}(2016){Woitke}, {Min}, {Pinte}, {Thi}, {Kamp}, {Rab}, {Anthonioz}, {Antonellini}, {Baldovin-Saavedra}, {Carmona}, {Dominik}, {Dionatos}, {Greaves}, {G{\"u}del}, {Ilee}, {Liebhart}, {M{\'e}nard}, {Rigon}, {Waters}, {Aresu}, {Meijerink}, \& {Spaans}}]{Woitke2016}
{Woitke}, P., {Min}, M., {Pinte}, C., {et~al.} 2016, \aap, 586, A103

\bibitem[{{Wuchterl} {et~al.}(2000){Wuchterl}, {Guillot}, \& {Lissauer}}]{Wuchterl2000}
{Wuchterl}, G., {Guillot}, T., \& {Lissauer}, J.~J. 2000, in Protostars and Planets IV, ed. V.~{Mannings}, A.~P. {Boss}, \& S.~S. {Russell}, 1081

\bibitem[{{Yang} {et~al.}(2017){Yang}, {Johansen}, \& {Carrera}}]{Yang2017}
{Yang}, C.-C., {Johansen}, A., \& {Carrera}, D. 2017, \aap, 606, A80

\bibitem[{{Yang} {et~al.}(2021){Yang}, {Sakai}, {Zhang}, {Murillo}, {Zhang}, {Higuchi}, {Zeng}, {L{\'o}pez-Sepulcre}, {Yamamoto}, {Lefloch}, {Bouvier}, {Ceccarelli}, {Hirota}, {Imai}, {Oya}, {Sakai}, \& {Watanabe}}]{Yang2021}
{Yang}, Y.-L., {Sakai}, N., {Zhang}, Y., {et~al.} 2021, \apj, 910, 20

\bibitem[{{Youdin} \& {Goodman}(2005)}]{Youdin2005}
{Youdin}, A.~N. \& {Goodman}, J. 2005, \apj, 620, 459

\bibitem[{{Zhang} {et~al.}(2015){Zhang}, {Blake}, \& {Bergin}}]{Zhang2015}
{Zhang}, K., {Blake}, G.~A., \& {Bergin}, E.~A. 2015, \apjl, 806, L7

\bibitem[{{Zhu} \& {Dong}(2021)}]{Zhu2021}
{Zhu}, W. \& {Dong}, S. 2021, \araa, 59, 291

\end{thebibliography}

\begin{appendix}
\section{Evolutionary model}
\label{app:model}

Our evolutionary model assumes that the central protostar accretes gas from its surrounding envelope with mass $M_{\rm env}(t)$ over time with a rate of $\dot{M}(t)$ to grow to a final protostellar mass of $M_{\star,\rm f}$. This accretion will result in an accretion luminosity of $L_{\rm acc}(t) = \eta G M_{\star}(t) \dot{M}/R_{\star}(t)$, where $\eta$ was assumed as 0.8 which is a typical value for young protostars and can be found by assuming that the disk accretes from an inner radius of ${\sim}5$\,R$_{\star}$ (\citealt{Meyer1997}; \citealt{Hartmann2016}). We prescribe the evolution of envelope mass following \cite{Fischer2017} as

\begin{equation}
    M_{\rm env}(t) = M_{\rm env,0} \times e^{-t\ln{2}/t_{\rm H}},
    \label{eq:M_env}
\end{equation}

\noindent where $M_{\rm env,0}$ is the initial envelope mass within a radius (assumed as 2500\,au here) and $t_{\rm H}$ is the time required for the protostar to accumulate half of its final mass. The accretion rate is defined simply by $M_{\rm env}/t_{\rm ff}$, where $t_{\rm ff}$ is the freefall time of the envelope within the radius which we defined $M_{\rm env,0}$ (i.e, 2500\,au). The protostellar mass ($M_{\star}$) would then increase with time using $\int_0^t{\dot{M}dt}$. The evolution of envelope and protostellar mass is shown in Fig. \ref{fig:evolution}a. We note that the final protostellar mass can be larger than the initial envelope mass within 2500\,au because the protostar can accrete mass from radii larger than 2500\,au.

For the protostellar luminosity, radius, and effective temperature (see Fig. \ref{fig:evolution}b and c) we first set the time at which the protostar has accumulated half of its mass ($t_{\rm H}$) as the birthline for the pre-main sequence evolutionary tracks of \cite{Baraffe2015}, and then take the protostellar parameters ($L_{\star}, T_{\star}$, and $R_{\star}$) for the corresponding protostellar mass at times $t-t_{\rm H}$ from \cite{Baraffe2015}. At times before the birthline, we take the protostellar parameters corresponding to the birthline of \cite{Baraffe2015} tracks while considering the evolving protostellar mass. We note that \cite{Baraffe2015} do not report $L_{\star}, T_{\star}$, and $R_{\star}$ at times below $5\times10^{5}$\,yr, thus we take their values at $5\times10^{5}$\,yr for the birthline and at times up to when $t - t_{\rm H}$ is equal to $5\times10^{5}$\,yr. This assumption is valid because \cite{Baraffe2002} showed that the protostellar luminosity does not change significantly at times before $5\times10^{5}$\,yr (see their Fig. 3). We also note that changing the birthline time from $t_{\rm H}$ to $t_{10\%}$ or to $t_{90\%}$ does not change the protostellar luminosity significantly (within ${\sim}25\%$). Finally, because the protostellar luminosity at times below $5\times10^{5}$\,yr is around one order of magnitude lower than the accretion luminosity (see Fig. \ref{fig:evolution}), these assumptions will not change the conclusions of this work.

\section{Full description of the grid parameters assumed for the radiative transfer}
\label{app:params}

The parameters used in the radiative transfer models for the cases where the final protostellar masses are 0.1, 0.5, and 1\,M$_{\odot}$ are given in Tables \ref{tab:params_0.1}, \ref{tab:params_0.5}, and \ref{tab:params_1}. Apart from the protostellar parameters which are prescribed as explained in Appendix \ref{app:model}, we opt for fixing the gas disk radius throughout the evolution (once at 50 and once at 100\,au). This is supported by current observations of gaseous disks (\citealt{Long2022}; \citealt{Hsieh2024}) and the chosen values are similar to typical dust disk radii found from \cite{Tobin2020} for their Class 0 objects when the dust and gas are more coupled than the later stages. However, we vary the disk mass during the evolution. We opt for a simple disk mass prescription. Based on the gaseous and dusty disk mass measurements (assuming a gas-to-dust mass ratio of 100) from millimeter and centimeter observations (\citealt{Miotello2016}; \citealt{Ansdell2016}; \citealt{Tychoniec2018,Tychoniec2020}; \citealt{Tobin2020}; \citealt{Anderson2024}), we assume gaseous disk masses of 0.02, 0.01, and 0.001\,M$_{\odot}$ for $t < 10^{5}$\,yr, $10^5\leq t < 10^{6}$\,yr, and $t \geq 10^6$\,yr, respectively. We note that based on planet formation models (e.g., \citealt{Savvidou2023}) a 0.02\,M$_{\odot}$ disk mass might not be enough to form a giant planet but at those early stages the disk and the planets could be constantly replenished by the envelope and these are approximate median masses, which disks may exceed by enough to grow a planet.

Following \cite{Nazari2022_disk}, we run the grid once for a dust distribution including large (maximum dust grain size of 1\,mm) grains and once for a dust distribution with small (0.1\,$\mu$m) grains. The reason to consider models with small and large grains is because at early times there is mixed evidence for dust growth in the inner envelope (\citealt{Ossenkopf1994}; \citealt{Kwon2009}; \citealt{Agurto2019}; \citealt{Cacciapuoti2023}; \citealt{Lebreuilly2023}), while high angular resolution observations reveal millimeter grain sizes in young disks that can hide the emission from various molecules (\citealt{Harsono2018}; \citealt{Nazari2024_wind}). We do not include dust drift and dust settling in the disks which is a reasonable assumption for young disks ($t{\lesssim}5\times10^5$\,yr, \citealt{Sheehan2022}; \citealt{Lin2023}; \citealt{vantHoff2023}; \citealt{Villenave2023}) but will become important at the later stages. Those effects are further discussed in Sect. \ref{sec:caveats}. We do not include viscous heating in the disk because $\dot{M}{\lesssim}10^{-5}$\,M$_{\odot}$\,yr$^{-1}$ at all evolutionary stages for the protostellar masses assumed here, in agreement with observations and models (e.g., \citealt{Alcala2017}; \citealt{Sellek2020}; \citealt{Fiorellino2023}; \citealt{Manara2023}; \citealt{Narang2024}). Thus the heating contribution from viscosity will be negligible (e.g., \citealt{Harsono2015}; \citealt{Nazari2023}).

\begin{table*}[t]
  \centering
  \small
  \caption{Model parameters and results for $M_{\star, \rm f} = 0.1$\,M$_{\odot}$}
  \label{tab:params_0.1}
  \setlength{\tabcolsep}{4pt}
  \begin{tabular}{lllllllllll}
  \toprule
  \toprule
  Parameter & $10^4$ & $5 \times 10^4$ & $10^5$ & $2 \times 10^5$& $3 \times 10^5$ & $5 \times 10^5$ & $10^6$ & $5 \times 10^6$ & $10^7$ & Description  \\
  & [yr] & [yr] &[yr] &[yr] &[yr] &[yr]&[yr] &[yr] &[yr] & \\
  \midrule
$L_{\star}$ [L$_{\odot}$]&      0.003&  0.02&   0.05&   0.07&   0.07&   0.08&   0.08&   0.03&   0.02&    Protostellar luminosity \\
$L_{\rm{acc}}$ [L$_{\odot}$]&   0.7&    1.0&    0.8&    0.3&    0.2&     $2 \times 10^{-2}$&     $2 \times 10^{-4}$&     0&    0&     Accretion luminosity \\
$T_{\star}$ [K]&        2388&   2731&   2834&   2861&   2925&   2945&   2952&   3049&   3074&    Protostellar effective temperature \\
$M_{\star}$ [M$_{\odot}$]&      0.01&   0.04&   0.07&   0.09&   0.10&   0.11&   0.11&   0.11&   0.11&    Protostellar mass\\
$M_{\rm env}$ [M$_{\odot}$]&    0.07&   0.05&   0.03&   0.01&     $4 \times 10^{-3}$&     $6 \times 10^{-4}$&     $4 \times 10^{-6}$&     0&    0&     Envelope mass in 2500\,au\\
$R_{\star}$ [R$_{\odot}$]&      0.3&    0.7&    0.9&    1.1&    1.0&    1.1&    1.1&    0.6&    0.4&     Protostellar radius\\
$\epsilon$ &    0.3&    0.3&    0.2&    0.2&    0.2&    0.2&    0.1&    0.1&    0.1&     Initial value of H/R\\
$M_{\rm disk}$ [M$_{\odot}$] &  0.02&   0.02&   0.01&   0.01&   0.01&   0.01&   0.001&  0.001&  0.001&   Disk mass\\
\midrule
$T_{\rm mid-plane}$ [K]&	343&	235&	112&	52&	43&	31&	35&	28&	25& Mid-plane temperature at 1\,au \\ 
$H/R$&	0.123&	0.101&	0.070&	0.048&	0.044&	0.037&	0.039&	0.035&	0.033& H/R at 1\,au \\ 
$M_{\rm gap}$ [M$_{\oplus}$]&	29&	17&	5&	2&	1&	0.8&	1.0&	0.7&	0.6& Min gap-opening planet mass at 1\,au \\
\midrule
$T_{\rm mid-plane}$ [K]&	58&	54&	53&	42&	35&	26&	30&	24&	22& Mid-plane temperature at 10\,au \\ 
$H/R$&	0.157&	0.152&	0.149&	0.134&	0.121&	0.105&	0.113&	0.101&	0.096& H/R at 10\,au \\ 
$M_{\rm gap}$ [M$_{\oplus}$]&	62&	56&	53&	38&	29&	19&	23&	17&	14& Min gap-opening planet mass at 10\,au \\ 
\bottomrule
  \end{tabular}
  \tablefoot{The parameters used for the radiative transfer models (top block) and the results of the models assuming $R_{\rm disk} = 50$\,au and small dust grains (bottom two blocks) for the case of $M_{\star, \rm f} = 0.1$\,M$_{\odot}$. Values below $10^{-10}$ are given as zero. Initial envelope mass, $M_{\rm env, 0}$, and the time required for the protostar to accrete half of its mass, $t_{\rm H}$, are taken as 0.079\,M$_{\odot}$ and $7\times 10^4$\,yr. For all models, the centrifugal radius ($R_{\rm c}$) was assumed as 50\,au, and the inner and outer radii of the grid are set to 0.4\,au and 150\,au. All models were run once with a disk radius ($R_{\rm disk}$) of 50\,au and once with $R_{\rm disk}$ of 100\,au.}
\end{table*}

\begin{table*}[t]
  \centering
  \small
  \caption{Model parameters and results for $M_{\star, \rm f} = 0.5$\,M$_{\odot}$}
  \label{tab:params_0.5}
  \setlength{\tabcolsep}{4pt}
  \begin{tabular}{lllllllllll}
  \toprule
  \toprule
  Parameter & $10^4$ & $5 \times 10^4$ & $10^5$ & $2 \times 10^5$ &$3 \times 10^5$ & $5 \times 10^5$ & $10^6$ & $5 \times 10^6$ & $10^7$ & Description  \\
  & [yr] & [yr] &[yr] &[yr] &[yr] &[yr]&[yr] &[yr] &[yr] & \\
  \midrule
    $L_{\star}$ [L$_{\odot}$]&      0.03&  0.29&   0.63&   0.91&   1.19&   1.19&   0.70&   0.20&   0.12&    Protostellar luminosity \\
    $L_{\rm{acc}}$ [L$_{\odot}$]&   6.2&    8.9&    7.1&    3.1&    1.1&    0.1&    $4 \times 10^{-4}$&     0&    0&     Accretion luminosity \\
    $T_{\star}$ [K]&        2761&   3220&   3460&   3672&   3849&   3849&   3807&   3697&   3651&    Protostellar effective temperature \\
    $M_{\star}$ [M$_{\odot}$]&      0.05&   0.19&   0.32&   0.44&   0.48&   0.50&   0.50&   0.50&   0.50&    Protostellar mass\\
    $M_{\rm env}$ [M$_{\odot}$]&    0.20&   0.12&   0.07&   0.02&   $7 \times 10^{-3}$&     $7 \times 10^{-4}$&     $2 \times 10^{-6}$&     0&    0&     Envelope mass in 2500\,au\\
    $R_{\star}$ [R$_{\odot}$]&      0.8&    1.7&    2.2&    2.4&    2.5&    2.5&    1.9&    1.1&    0.9&     Protostellar radius\\
    $\epsilon$ &    0.3&    0.3&    0.2&    0.2&    0.2&    0.2&    0.1&    0.1&    0.1&     Initial value of H/R\\
    $M_{\rm disk}$ [M$_{\odot}$] &  0.02&   0.02&   0.01&   0.01&   0.01&   0.01&   0.001&  0.001&  0.001&   Disk mass\\
    \midrule
    $T_{\rm mid-plane}$ [K]&	389&	293&	193&	105&	66&	54&	36&	29&	27& Mid-plane temperature at 1\,au \\ 
$H/R$&	0.058&	0.051&	0.041&	0.030&	0.024&	0.022&	0.018&	0.016&	0.015& H/R at 1\,au \\ 
$M_{\rm gap}$ [M$_{\oplus}$]&	16&	10&	6&	2&	1&	0.9&	0.5&	0.3&	0.3& Min gap-opening planet mass at 1\,au \\ 
    \midrule
    $T_{\rm mid-plane}$ [K]&	90&	99&	97&	75&	57&	39&	41&	32&	29& Mid-plane temperature at 10\,au \\ 
$H/R$&	0.087&	0.092&	0.091&	0.080&	0.069&	0.057&	0.059&	0.052&	0.050& H/R at 10\,au \\ 
$M_{\rm gap}$ [M$_{\oplus}$]&	53&	61&	60&	41&	27&	15&	17&	11&	10& Min gap-opening planet mass at 10\,au \\ 
    \bottomrule
  \end{tabular}
  \tablefoot{The parameters used for the radiative transfer models (top block) and the results of the models assuming $R_{\rm disk} = 50$\,au and small dust grains (bottom two blocks) for the case of $M_{\star, \rm f} = 0.5$\,M$_{\odot}$. Values below $10^{-10}$ are given as zero. Initial envelope mass, $M_{\rm env, 0}$, and the time required for the protostar to accrete half of its mass, $t_{\rm H}$, are taken as 0.219\,M$_{\odot}$ and $6\times 10^4$\,yr. For all models, the centrifugal radius ($R_{\rm c}$) was assumed as 50\,au, and the inner and outer radii of the grid are set to 0.4\,au and 150\,au. All models were run once with a disk radius ($R_{\rm disk}$) of 50\,au and once with $R_{\rm disk}$ of 100\,au.}
\end{table*}

\begin{table*}[t]
  \centering
  \small
  \caption{Model parameters and results for $M_{\star, \rm f} = 1$\,M$_{\odot}$}
  \label{tab:params_1}
  \setlength{\tabcolsep}{4pt}
  \begin{tabular}{lllllllllll}
  \toprule
  \toprule
  Parameter & $10^4$ & $5 \times 10^4$ & $10^5$ & $2 \times 10^5$& $3 \times 10^5$& $5 \times 10^5$ & $10^6$ & $5 \times 10^6$ & $10^7$ & Description  \\
  & [yr] & [yr] &[yr] &[yr] &[yr] &[yr]&[yr] &[yr] &[yr] & \\
  \midrule
$L_{\star}$ [L$_{\odot}$]&      0.08&  1.19&   1.85&   3.22&   3.22&   3.22&   2.02&   0.65&   0.46&    Protostellar luminosity \\
$L_{\rm{acc}}$ [L$_{\odot}$]&   24.8&   33.2&   27.4&   8.5&    2.3&    0.1&    $2 \times 10^{-4}$&     0&    0&     Accretion luminosity \\
$T_{\star}$ [K]&        2945&   3849&   4111&   4397&   4397&   4397&   4379&   4310&   4363&    Protostellar effective temperature \\
$M_{\star}$ [M$_{\odot}$]&      0.11&   0.46&   0.73&   0.96&   1.02&   1.04&   1.04&   1.04&   1.04&    Protostellar mass\\
$M_{\rm env}$ [M$_{\odot}$]&    0.34&   0.19&   0.10&   0.02&   $6 \times 10^{-3}$&     $4 \times 10^{-4}$&     $4 \times 10^{-7}$&     0&    0&     Envelope mass in 2500\,au\\
$R_{\star}$ [R$_{\odot}$]&      1.1&    2.5&    2.7&    3.1&    3.1&    3.1&    2.5&    1.4&    1.2&     Protostellar radius\\
$\epsilon$ &    0.3&    0.3&    0.2&    0.2&    0.2&    0.2&    0.1&    0.1&    0.1&     Initial value of H/R\\
$M_{\rm disk}$ [M$_{\odot}$] &  0.02&   0.02&   0.01&   0.01&   0.01&   0.01&   0.001&  0.001&  0.001&   Disk mass\\
\midrule
$T_{\rm mid-plane}$ [K]&	436&	366&	275&	139&	71&	38&	40&	38&	31& Mid-plane temperature at 1\,au \\ 
$H/R$&	0.044&	0.040&	0.035&	0.025&	0.017&	0.011&	0.013&	0.013&	0.012& H/R at 1\,au \\ 
$M_{\rm gap}$ [M$_{\oplus}$]&	13&	10&	7&	2&	1&	0.5&	0.4&	0.3&	0.3& Min gap-opening planet mass at 1\,au \\ 
\midrule
$T_{\rm mid-plane}$ [K]&	140&	166&	156&	100&	70&	43&	48&	38&	36& Mid-plane temperature at 10\,au \\ 
$H/R$&	0.077&	0.084&	0.081&	0.065&	0.054&	0.043&	0.045&	0.040&	0.039& H/R at 10\,au \\ 
$M_{\rm gap}$ [M$_{\oplus}$]&	73&	94&	85&	44&	25&	12&	15&	10&	9& Min gap-opening planet mass at 10\,au \\ 
\bottomrule
  \end{tabular}
  \tablefoot{The parameters used for the radiative transfer models (top block) and the results of the models assuming $R_{\rm disk} = 50$\,au and small dust grains (bottom two blocks) for the case of $M_{\star, \rm f} = 1$\,M$_{\odot}$. Values below $10^{-10}$ are given as zero. Initial envelope mass, $M_{\rm env, 0}$, and the time required for the protostar to accrete half of its mass, $t_{\rm H}$, are taken as 0.39\,M$_{\odot}$ and $5\times 10^4$\,yr. For all models, the centrifugal radius ($R_{\rm c}$) was assumed as 50\,au, and the inner and outer radii of the grid are set to 0.4\,au and 150\,au. All models were run once with a disk radius ($R_{\rm disk}$) of 50\,au and once with $R_{\rm disk}$ of 100\,au.}
\end{table*}

\section{Results of the models with large grains and $R_{\rm disk}=100$\,au}
\label{app:other_models}

\begin{figure*}
    \centering
    \includegraphics[width=0.9\textwidth]{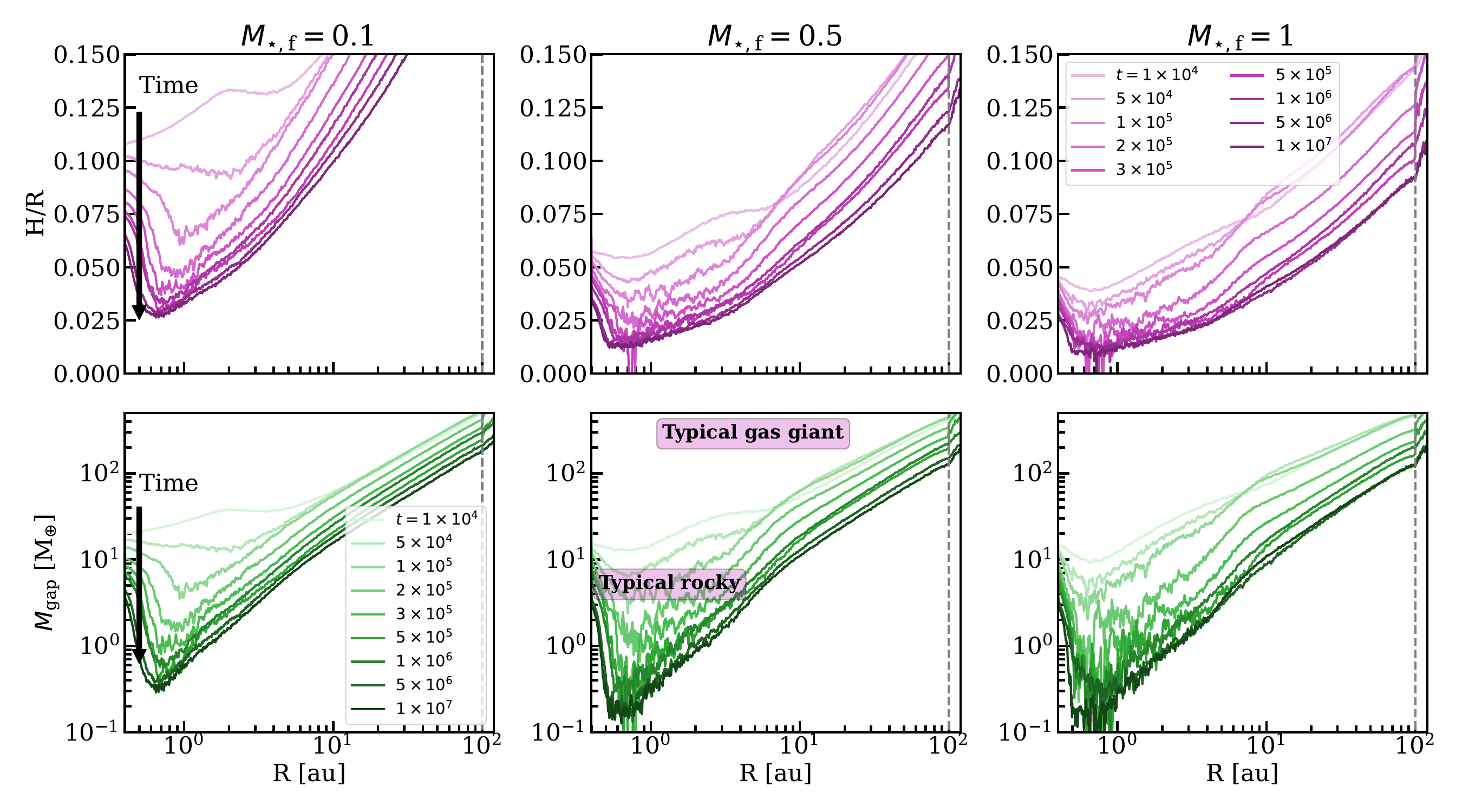}
    \caption{Same as Fig. \ref{fig:min_pl_mass} but for models with disk radius of 100\,au.} 
    \label{fig:min_pl_mass_100au}
\end{figure*}

\begin{figure*}
    \centering
    \includegraphics[width=0.9\textwidth]{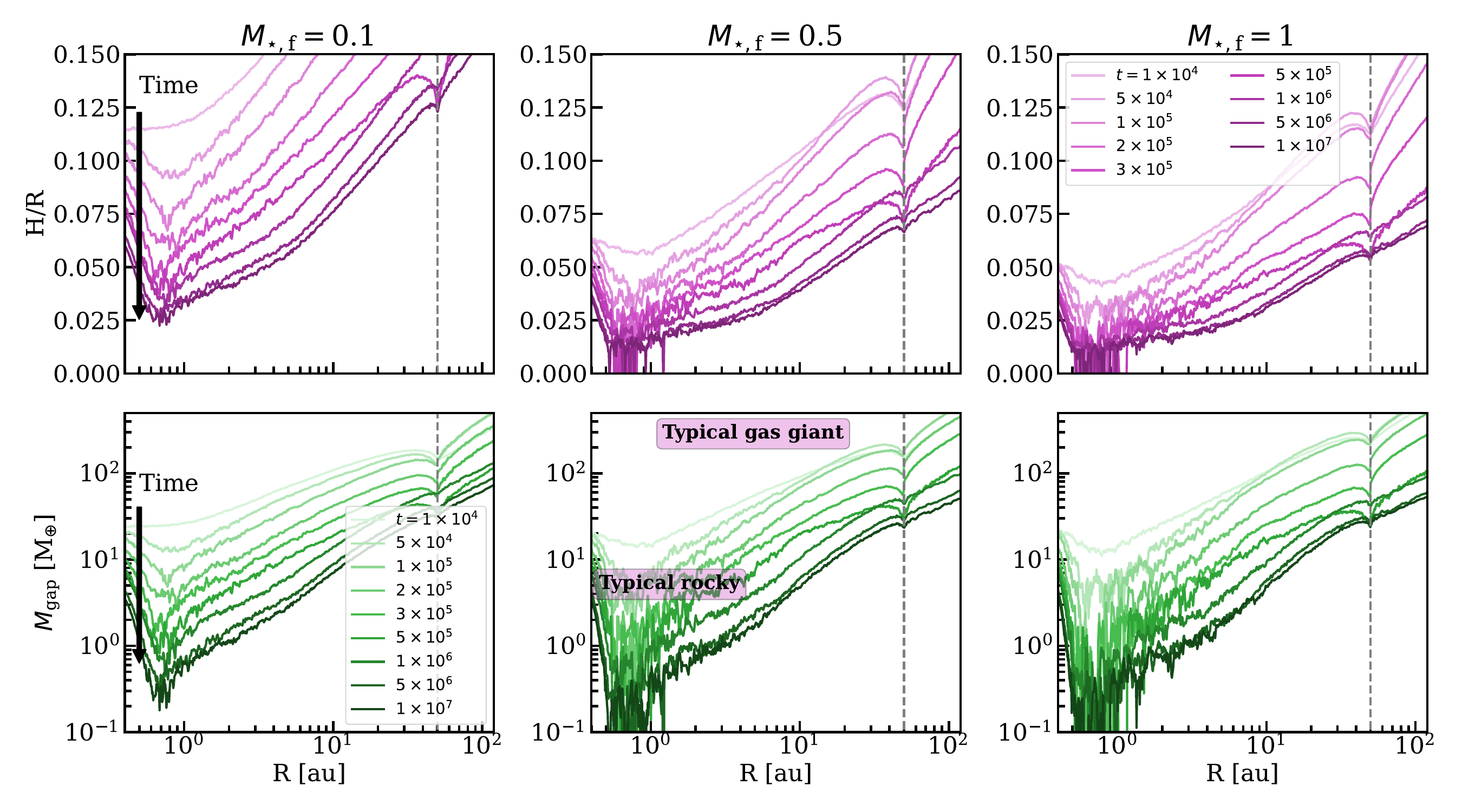}
    \caption{Same as Fig. \ref{fig:min_pl_mass} but for models with large dust grains.} 
    \label{fig:min_pl_mass_large}
\end{figure*}

Figure \ref{fig:min_pl_mass_100au} presents the disk aspect ratio and minimum gap-opening planet mass for the models with disk radius of 100\,au, while Fig. \ref{fig:min_pl_mass_large} shows those for models with large dust grains (maximum grain size of 1\,mm) and $R_{\rm disk}=50$\,au. The models with a larger disk radius show similar trends and values for the disk aspect ratio and minimum gap-opening planet mass at radii below 50\,au to the models with $R_{\rm disk} = 50$\,au. Moreover, including large grains in the models does not change the trend in aspect ratio with time (see Fig. \ref{fig:min_pl_mass_large}). Including large grains in some models and at certain radii increases the aspect ratio and in some other cases it decreases or plays no significant role in the disk aspect ratio. Nevertheless, the changes produced in aspect ratio by including larger grains are for most cases below 0.025.

\section{Gap-opening planet mass as a function of time}
\label{app:function}

In the left panel of Fig. \ref{fig:min_pl_mass_time} we fit a line to the findings of the radiative transfer models for the case where $M_{\star, \rm f} = 0.5$\,M$_{\odot}$ as shown by the gray dashed line. We use \texttt{curve\_fit} from \texttt{scipy.optimize} in \texttt{python}. The function that we use to fit the models with takes the following form

\begin{equation}
    M_{\rm gap}(t) = \frac{a + b (t/10^6\,\mathrm{yr})^d \times e^{\log_{10}(t/t_0)/c}}{ 1.0 + e^{\log_{10}(t/t_0)/c}},
    \label{eq:fitted}
\end{equation}

\noindent with the final values of $a, b, c, d$, and $t_0$ given by 58.2\,M$_{\oplus}$, 14.9\,M$_{\oplus}$, 0.09, -0.17, and $10^{5.33}$\,yr respectively.

\section{Pebble Accretion}
\subsection{Toy Model}
\label{sec:pebble_accretion}
In order to estimate for how many disks the growth timescale can compete within the lifetime of the embedded phase (and thus elevated temperatures), we use a simple pebble accretion model based on \citet{Lambrechts2014b}. Following that work we use a coagulation efficiency of 0.5 everywhere.
Here the pebble accretion rate is assumed to be given by the 2D Hill regime:
\begin{equation}
    \dot{M}_{\rm{c}} = 2 \left(\frac{S\!t}{0.1}\right)^{2/3} r_{\rm{H}}^2 \Omega_{\rm K} \Sigma_{\rm peb}
    \label{eq:2DHill}
    ,
\end{equation}
where $r_H$ and $\Omega_{\rm K}$ are the Hill radius and Keplerian orbital frequency respectively, the Stokes number is assumed to be given by the drift limit and depends on the gas pressure ($P$) gradient and gas surface density ($\Sigma_{\rm g}$)
\begin{equation}
    S\!t = \frac{\sqrt{3}}{8} \left(H/R\right)^{-2} \left(\frac{\partial \ln P}{\partial \ln R}\right)^{-1} \frac{\Sigma_{\rm peb}}{\Sigma_{\rm g}}
    \label{eq:Stdrift}
    ,
\end{equation}
and the steady-state pebble surface density is
\begin{equation}
    \Sigma_{\rm peb} = \sqrt{\frac{4}{\pi\sqrt{3}}\frac{\dot{M}_{\rm peb}\Sigma_{\rm g}}{R^2 \Omega_{\rm K}}}
    \label{eq:Sigmapeb}
    .
\end{equation}

We assume that the pebbles are produced due to the progress of a growth front outwards through a disk with gas surface density $\Sigma_{\rm g} = \Sigma_1 (R/\rm{au})^{-1}$ (where $\Sigma_1$ is the gas surface density at 1 au, called $\beta$ by \citealt{Lambrechts2014b}) and initial dust-to-gas ratio $Z_0=0.01$. Then following \citet{Lambrechts2014b}, the pebble flux is
\begin{equation}
    \dot{M}_{\rm peb} = (1/3)^{2/3} \pi (GM_*\, \mathrm{au}^3)^{1/3} \Sigma_1 Z_0^{5/3} t^{-1/3} f_{\rm PF} 
    ,
\end{equation}
where we introduce the factor $f_{\rm PF}$ to allow for a suppression of the pebble fluxes (see main text).

\begin{figure*}
    \centering
    \includegraphics[width=0.9\textwidth]{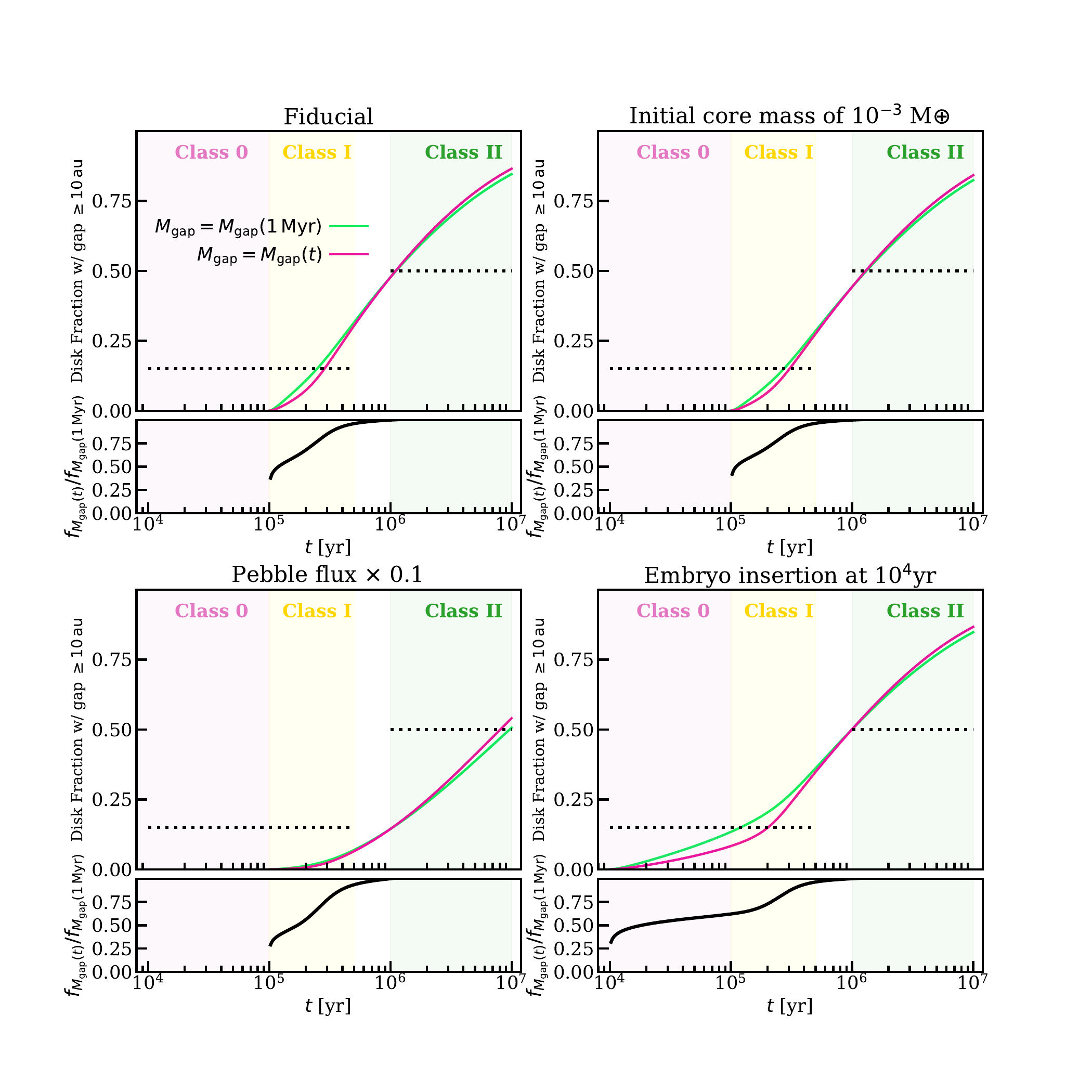}
    \caption{Same as right column of Fig. \ref{fig:min_pl_mass_time} (top left here), showing the effect of varying several parameters that are inputted in our toy model. The top right assumes a smaller initial embryo mass, bottom left presents the case where the pebble flux is decreased by a factor of 10 and bottom right panel shows what happens if the embryo is inserted in the Class 0 phase. Although the absolute value of the disk fractions with a gap at ${\geq}10$\,au varies significantly depending on our assumptions, the difference between the case where we consider consider the temperature effects and the case where we do not (the green and pink lines) remains a factor of ${\gtrsim}2$.}
    \label{fig:multi}
\end{figure*}

Combining everything, we obtain a pebble accretion rate

\begin{align}
    \dot{M}_{\rm{c}} =& 4.8\,M_{\oplus}\,\mathrm{Myr^{-1}} 
    f_{\rm PF}^{5/6}
    \left(\frac{Z_0}{0.01}\right)^{25/18} \left(\frac{M_*}{M_{\odot}}\right)^{-11/36} \left(\frac{M_{\rm c}}{M_{\oplus}}\right)^{2/3}  
    \nonumber \\
    &\times \left(\frac{\Sigma_1}{500\,\mathrm{g\,cm^{-2}}}\right) \left(\frac{R}{10\,\mathrm{au}}\right)^{-5/12} \left(\frac{(H/R)_t}{0.059}\right)^{-4/3} \left(\frac{t}{\mathrm{Myr}}\right)^{-5/18}.
    \label{eq:Mdot_core}
\end{align}
Compared to Eq. 31 of \citet{Lambrechts2014b}, our Eq. \ref{eq:Mdot_core} has the additional suppression factor $f_{\rm PF}^{5/6}$.
To test whether these effects could change our conclusions, we also investigate a scenario where the pebble fluxes are reduced by a factor $f_{\rm PF}=0.1$ (see Fig. \ref{fig:multi}), i.e. only 10 \% of the dust grains produced at the growth front are assumed to be efficiently drifting pebbles.

Equation \ref{eq:Mdot_core} also has an additional $(H/R)^{-4/3}$ term which results from the larger absolute pressure gradient in younger, hotter, disks which lowers the Stokes number in the drift limit (Eq. \ref{eq:Stdrift}) and thus reduces the efficiency of pebble accretion (Eq. \ref{eq:2DHill}) in young, more vertically-extended disks. It is normalized to $(H/R)=0.059$ at 10 au, which is the value appropriate to \cite{Lambrechts2014b}, and also happens to be the value at 1 Myr for our $0.5$\,M$_{\odot}$ models (Table \ref{tab:params_0.5}).

Although the factors of the planet location $R$ and background density $\Sigma_1$ may change over time due to planet migration or disk dispersal respectively, we hold them constant here both for simplicity and to better isolate the effects of the changing temperatures.
We may then integrate Eq. \ref{eq:Mdot_core} and solve for the required $\Sigma_1$ to grow a planet of mass $M_{\rm gap}(t)$ in time t:
\begin{align}
    \frac{\Sigma_{1, \rm crit}(t)}{500\,\mathrm{g\,cm^{-2}}} =& \frac{11^{-1/3}}{f_{\rm PF}^{5/6}} \left(\frac{Z_0}{0.01}\right)^{-25/18} \left(\frac{M_*}{M_{\odot}}\right)^{11/36} \left(\frac{R}{10\,\mathrm{au}}\right)^{5/12}
    \nonumber\\
    &\times \frac{M_{\rm gap}(t)^{1/3}-M_{\rm c,0}^{1/3}}{\frac{13}{18} \int_{t_{\rm i}}^{t}{ t^{-5/18} \left(\frac{(H/R)_t}{0.059}\right)^{-4/3} {\rm d}t}},
    \label{eq:S1_crit}
\end{align}
where the integral is performed numerically and $H/R$ changes with time as a function of the same form as Eq. \ref{eq:fitted}. 

Finally, we convert $\Sigma_{1, \rm crit}$ to a dust mass by assuming the same 1/R profile and a typical disk radius of $50\,\mathrm{au}$ \citep{Tobin2020}
\begin{equation}
    M_{\rm d, crit}(t) = 100\pi\,\mathrm{au}^2 Z_0 \Sigma_{1, \rm crit}(t).
    \label{eq:Md_crit}
\end{equation}
Following \citet{Tobin2020} we assume the disk mass follows a lognormal distribution with median $\bar{M}_{\rm d}=52.5\,M_{\oplus}$ and $\sigma=0.83$ such that we can estimate the fraction of disks which can grow a gap-opening planet in time $t$ using the error function:
\begin{equation}
    f_{M_{\rm gap}(t)} = \frac{1}{2} \left( 1-{\rm erf}\left[\frac{\log_{10}(M_{\rm d, crit}(t)/\bar{M}_{\rm d})}{\sqrt{2}\sigma}\right] \right)
    \label{eq:fgap_t}
    .
\end{equation}

\noindent This equation is used to produce the right panels of Fig. \ref{fig:min_pl_mass_time}. As well as the fiducial values presented in Sect. \ref{sec:implications} and Fig. \ref{fig:min_pl_mass_time}, in Fig. \ref{fig:multi}, we also present results for a lower core mass of $M_{\rm c,0}=10^{-3}\,M_{\oplus}$, earlier insertion time $t_{\rm i}=0.01\,\mathrm{Myr}$ and reduced pebble flux $f_{\rm PF}=0.1$, demonstrating that the main conclusion is largely insensitive to these parameters. More information on the consequences of Eq. \ref{eq:fgap_t} are given in Sect. \ref{sec:implications}.

To contextualize these numbers, we note that at the end of the Class I phase $M_{\rm crit} = 108\,M_{\oplus} \approx 2 \bar{M}_{\rm d}$ and $M_{\rm gap}(t)=12.2\,M_{\oplus}$. This corresponds to an average $\sim11$ \% pebble accretion efficiency, consistent with expectations \citet{Lambrechts2014b}, assuming the whole initial dust mass can participate. This is not a bad assumption, the remaining disk mass estimated from ALMA fluxes in the Class I phase is already $\sim 25\%$ of the Class 0 mass, falling to $\sim5\%$ by Class II \citep{Tychoniec2020} suggesting significant dust loss to pebble drift. Nevertheless, we note that our toy model is likely too simple to truly capture planet formation at early times and thus may be overestimating the absolute number of gaps that can be produced by the growing planets.

\subsection{3D Pebble Accretion Regime}
\label{sec:pebble_accretion_3D}
Lower-mass embryos can have accretion radii that are smaller than the pebble scale height, in which case they will accrete in the 3D regime, which is less efficient since they can only access a fraction of the disc's pebbles.
Since the dust scale height may be written in terms of the turbulence parameter $\alpha$ as $H_{\rm{d}}/H = \sqrt{\frac{\alpha}{S\!t+\alpha}}$ then for pebbles (where $S\!t\gg\alpha$) $H_{\rm peb}/H \sim \sqrt{\frac{\alpha}{S\!t}}$. Following Eqs. 10 and Fig. 7 of \cite{Drkazkowska2023}, we may estimate $\dot{M}_{\rm{c,3D}}/\dot{M}_{\rm{c,2D}} \approx \left(\frac{S\!t}{0.2}\right)^{5/6} \left(\frac{\alpha}{10^{-4}}\right)^{-1/2} \left( \frac{M_{\rm c}/M_{*}}{3\times10^{-9}} \right)$. For $\alpha=10^{-4}-10^{-3}$, a $10^{-3}\,M_{\oplus}$ embryo \citep[as used by][]{Lambrechts2014b} accretes in the 2D Hill regime for Stokes numbers of $S\!t=0.1-1$. For the median Class 0 disk mass \citep{Tobin2020}, such Stokes numbers corresponds to $\sim$cm-sized pebbles in the outer disk at ($\sim$50 au). The corresponding numbers for a $10^{-1}\,M_{\oplus}$ embryo are $S\!t\approx10^{-2}$ and $\sim$mm-sized pebbles.

The equivalents of Equations \ref{eq:2DHill},  \ref{eq:Mdot_core} and \ref{eq:S1_crit} in the 3D regime are
\begin{equation}
    \dot{M}_{\rm{c}} = 0.12 \left(\frac{S\!t}{0.1}\right)^{3/2} \frac{r_{\rm{H}}^3}{H\sqrt{\alpha}} \Omega_{\rm K} \Sigma_{\rm peb}
    \label{eq:3DBondi}
    ,
\end{equation}
\begin{align}
    \dot{M}_{\rm{c}} =& 1.7\,M_{\oplus}\,\mathrm{Myr^{-1}} 
    f_{\rm PF}^{5/4} \left(\frac{\alpha}{10^{-3}}\right)^{-1/2}
    \left(\frac{Z_0}{0.01}\right)^{25/12} \left(\frac{M_*}{M_{\odot}}\right)^{-5/18} \left(\frac{M_{\rm c}}{M_{\oplus}}\right)  
    \nonumber \\
    &\times \left(\frac{\Sigma_1}{500\,\mathrm{g\,cm^{-2}}}\right) \left(\frac{R}{10\,\mathrm{au}}\right)^{-5/24} \left(\frac{(H/R)_t}{0.059}\right)^{-19/6} \left(\frac{t}{\mathrm{Myr}}\right)^{-5/12}
    \label{eq:Mdot_core_3D}
    ,
\end{align}
\begin{align}
    \frac{\Sigma_{1, \rm crit}(t)}{500\,\mathrm{g\,cm^{-2}}} =& \frac{0.12}{f_{\rm PF}^{5/4}} \left(\frac{Z_0}{0.01}\right)^{-25/12} \left(\frac{M_*}{M_{\odot}}\right)^{5/18} \left(\frac{R}{10\,\mathrm{au}}\right)^{5/24}
    \nonumber\\
    &\times \frac{\ln(M_{\rm c,3-2}/M_{\rm c,0})}{\frac{7}{12} \int_{t_{\rm i}}^{t}{ t^{-5/12} \left(\frac{(H/R)_t}{0.059}\right)^{-19/6} {\rm d}t}}
    \label{eq:S1_crit_3D}
    .
\end{align}
where $M_{\rm c,3-2}$ is the core mass at the transition from 3D to 2D pebble accretion.

\begin{figure}
    \centering
    \includegraphics[width=\linewidth]{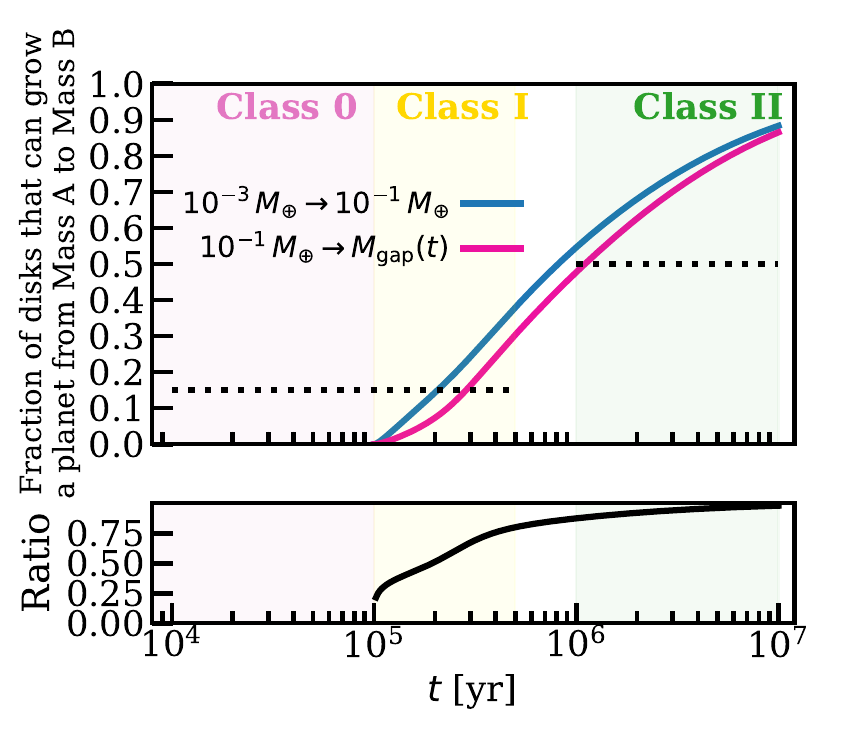}
    \caption{Comparison between the fraction of disks which can grow planets through the 3D regime (here estimated as $10^{-3}-10^{-1}\,M_{\oplus}$) in blue and those that can grow a planet to the gap-opening mass in the 2D regime (here estimated as $>10^{-1}\,M_{\oplus}$) in pink (which is the same as the pink line in Fig. \ref{fig:min_pl_mass_time}.)} 
    \label{fig:3Dvs2D}
\end{figure}

From these we can, similarly to above, estimate the fraction of disks that can grow a planet through the 3D regime in time $t$. In Figure \ref{fig:3Dvs2D} we show that it is always greater than the fraction that can grow a gap-opening planet in the fiducial case\footnote{This results from the different mass ranges considered; if the 2D rates were used for the same mass range they would produce even higher fractions.}; thus in our toy model planets are expected to spend less time in the 3D regime and it should not be the limiting factor.

\section{Gravitational Instability}
\label{sec:grav_instabil}
Our discussion in Section \ref{sec:discuss} focused on planet formation by core accretion, starting with pebble accretion. One major potential alternative planet-forming mechanism is fragmentation due to gravitational instability (GI).
A common criterion for instability is that the disk becomes unstable when the disc-to-star mass ratio is greater than the aspect ratio. Given the aspect ratios in Tables \ref{tab:params_0.1}-\ref{tab:params_1}, ${\sim}20\%$ of disks from the \citet{Tobin2020} Class 0 non-multiple mass distribution may be massive enough to be gravitationally unstable. However, fragmentation may not be the outcome of the GI instabilities if they can instead efficiently transport angular momentum and regulate the disk mass through accretion \citep{Kratter2008}. High levels of irradiation - as found in the Class 0 phase - may have both stabilizing and destabilizing effects against fragmentation \citep{Rice2011} so more dedicated modeling would be needed to deduce the fraction of embedded disks that can fragment. Nevertheless, in such young disks, fragmentation is more likely to form more massive companions on wide orbits \citep[unless dust drag acts to lower the length/mass-scale on which the instability operates][]{Longarini2023_theory,Longarini2023_sim,Baehr2023}, although the wider orbit planetary companions may migrate inward. Indeed ${~\sim}10-30\%$ of the VANDAM survey Class 0 disks host companions at $20-500\,\mathrm{au}$, which may indicate a disk-fragmentation formation route distinct from the turbulent cloud fragmentation population at 1000s au \citep{Tobin2022}. Such massive companions are also likely to drive spiral density waves as well as opening gaps; spiral patterns can also form in GI turbulent disks \citep{Kratter2016}.

\end{appendix}

\end{document}